\documentclass[preprint,3p,12pt,times]{elsarticle}

\usepackage{graphicx}
\usepackage{epsfig}
\usepackage{epstopdf}
\usepackage{amssymb}
\usepackage{amsthm}
\usepackage{bm}
\usepackage{placeins}

\usepackage{algpseudocode}

\usepackage[colorlinks=true,citecolor=blue,linkcolor=black]{hyperref}
\usepackage{color}
\usepackage{amsmath}
\usepackage{multirow}
\usepackage[justification=centering]{caption}
\usepackage{subcaption}
\usepackage{color}
\usepackage[algoruled,boxed,lined]{algorithm2e}
\newcommand{\norm}[1]{\left\lVert#1\right\rVert}

\usepackage{booktabs} 
\usepackage{lineno}

\journal{Elsevier}

\begin{document}

\title{Three-dimensional SPH modeling of brittle fracture under hydrodynamic loading}

\author[iitd]{Vishabjeet Singh}
\ead{amz238675@am.iitd.ac.in}
\author[lab,seu]{Chong Peng\corref{coraut}}
\ead{chong.peng@seu.edu.cn} 
\author[iitd]{Md Rushdie Ibne Islam\corref{coraut}}
\ead{rushdie@am.iitd.ac.in}

\cortext[coraut]{Corresponding Author}
\address[iitd]{Department of Applied Mechanics, Indian Institute of Technology Delhi, New Delhi 110016, India}
\address[lab]{State Key Laboratory of Safety, Durability and Healthy Operation of Long Span Bridges, Southeast University, Nanjing 211189, China}
\address[seu]{School of Civil Engineering, Southeast University, Nanjing 211189, China}

\begin{abstract}
A three-dimensional SPH computational framework is presented for modeling fluid-structure interactions with structural deformation and failure. We combine weakly compressible SPH with a pseudo-spring-based SPH solver to capture the fluid flow and deformable structures. A unified modeling approach captures the solid boundaries and fluid-structure interfaces without penalty-based contact force. The $\delta$-SPH technique improves the pressure calculations in the fluid phase, while structural damage is modeled using a pseudo-spring approach, with particle interactions limited to its neighbors. The present framework can capture the three-dimensional crack surfaces in structures without any computationally intensive crack-tracking algorithm or visibility criteria. The framework has been proven effective against existing models and experimental data, demonstrating high accuracy and robustness in simulating detailed fracture patterns and offering insights into the impact of hydrodynamic events on structural integrity.
\end{abstract}


\maketitle

\section{Introduction}
Understanding fluid-structure interaction (FSI) is crucial for designing safe industrial and engineering systems like ships, airplanes, automobiles, offshore platforms, and bridges because the interaction dominates the structures' stress and deformation behaviors. FSI is mainly studied through experimental, analytical, and numerical approaches. Recent advancements in numerical methods and high-performance computing have made numerical methods more and more applicable for addressing complex problems in FSI. 

Conventional mesh-based numerical methods, such as finite difference method (FDM), finite volume method (FVM), and finite element method (FEM) \cite{slone2002dynamic, heil2004efficient, hu2009two} are employed in FSI simulations; however, these methods struggle with handling dynamic interfaces, including free surfaces and deformable structures, often necessitating additional approaches like interface tracking or re-meshing. Such additional approaches demand significant implementation and computation effort and can compromise stability and accuracy. Moreover, many FSI problems have large deformations and crack propagations in structures, which are also inconvenient to simulate with mesh-based methods. Lagrangian particle-based methods offer an alternative, which can manage complex dynamic boundaries/interfaces, large deformations, and crack propagation. This versatility makes them ideal for FSI simulations.

Smoothed Particle Hydrodynamics (SPH) is a versatile computational approach introduced for modeling astrophysical phenomena \cite{gingold1977smoothed, lucy1977numerical}. Over time, it has gained popularity in solid and fluid mechanics fields owing to its ability to handle complex physical problems without grids. Instead of a grid, SPH employs discrete particles to represent the continuum \cite{libersky2005smooth, liu2010smoothed}, making it well-suited for scenarios involving large deformations, free surfaces, moving boundaries, and crack propagation \cite{monaghan1994simulating, adami2012generalized, fraga2019implementation}. SPH is especially effective in simulating free-surface flows and fluid-structure interactions (FSI). Its mesh-free approach simplifies the handling of intricate geometries without re-meshing, offering greater flexibility. However, SPH can be computationally demanding. With the advancement of hardware, SPH is expected to have wider applications.

SPH methods for fluid-structure interaction (FSI) \cite{antoci2007numerical, rafiee2009sph, salehizadeh2022coupled, zhan2019stabilized, islam2024sph} are categorized by the different techniques used for fluids and solids. For fluid dynamics, weakly compressible SPH (WCSPH) and incompressible SPH (ISPH) are commonly employed \cite{marrone2015prediction, meringolo2017filtering, pahar2016modeling, pahar2017coupled}, while for solids, conventional Eulerian SPH (ESPH) and total-Lagrangian SPH (TLSPH) are used. TLSPH improves stability and accuracy by using Lagrangian coordinates and kernel gradient correction, respectively. Additionally, parallel computing, particularly with GPU acceleration, has significantly enhanced SPH's computational efficiency \cite{zhan2019stabilized,dominguez2022dualsphysics}, which makes three-dimensional large-scale SPH simulations of FSI possible. Despite these advances, challenges remain in stability, accuracy, and computational efficiency, particularly for large-scale 3D FSI simulations. WCSPH can suffer from pressure-related inaccuracies, leading to efforts like $\delta$-SPH \cite{molteni2009simple, marrone2011delta} and low-dissipation Riemann solvers to improve pressure fields \cite{zhang2017weakly}. For solid modeling, issues like inconsistency and hourglass modes have led to improvements such as kernel gradient correction and artificial stress techniques \cite{monaghan2000sph, gray2001sph}. In this regard, a two-dimensional implicit SPH-based FSI solver is proposed for improved accuracy, consistency, and robustness in FSI scenarios \cite{shimizu2022implicit}. A coupled solver based on Riemann SPH for fluids and Hamiltonian SPH for structures is developed in \cite{khayyer2024improved} to improve volume conservation, incompressibility, and stability. A variationally consistent total Lagrangian SPH model with second-order completeness and a novel Riemann-based stabilisation term accounting for volumetric and shear wave effects is proposed in \cite{gotoh2025enhanced}. This framework provides an accurate and robust simulation of nonlinear structural dynamics under large deformations. An FSI solver based on total Lagrangian SPH with a parameter-free Dynamic Hourglass Control (DHGC) scheme is utilised in \cite{zhan2025enhanced} to suppress spurious zero-energy modes by dynamically adjusting the control coefficient based on particle distribution and error propagation. TLSPH, while more accurate, struggles with material distortion due to negative Jacobians \cite{islam2023comparison, islam2022large}.

Modeling of cracks in deformable structures under hydrodynamic loading remains a challenging task. An integrated computational framework combining the FEM, the Particle Finite Element Method (PFEM), and the Discrete Element Method (DEM) for simulating the failure of reinforced concrete structures subjected to impulsive wave forces from free-surface flows is proposed in \cite{onate2022combination}. In this approach, PFEM is employed to model the fluid flow, while structural response and fracture are captured using a coupled FEM–DEM formulation. The failure of quasi-brittle materials such as concrete and ice subjected to high-velocity water jet impacts are modeled using FEM-SPH \cite{yu2021sph} and Peridynamics-SPH \cite{shi2024quasi} coupled frameworks. A phase-field model and coupled peridynamics-SPH frameworks have been developed \cite{sun2020smoothed, liu2020coupled, dai2023coupled} to capture the structural failure subjected to different hydrodynamic loadings. An enhanced brittle fracture model based on the meshless Lattice Particle Method (LPM) coupled with SPH is presented in \cite{ng2023numerical}. The framework is implemented within the open-source DualSPHysics code, optimized for both CPU and GPU architectures. A coupled phase-field model in SPH setup is employed to simulate brittle fracture, capturing damage and failure in the deformable solid structure \cite{rahimi2023sph}. Recently, a pseudo-spring-ESPH \cite{chakraborty2013pseudo, islam2017computational} coupled with WCSPH in 2-D has been proposed to model crack propagation in deformable structures subjected to hydrodynamic loading \cite{islam2024sph}. The interaction between the fluid and the deformable structures is modeled by a soft repulsive particle contact model, which requires calibration of contact force. This process is challenging since each problem demands a unique set of contact parameters. The pseudo-spring analogy \cite{chakraborty2013pseudo, islam2017computational} is used to model the crack initiation, propagation and fracture. This approach models interactions between the immediate neighbouring particles using pseudo-springs. Crack propagation occurs when these pseudo-springs rupture, allowing the fracture to advance through the material. 

In this paper, we systematically developed a 3-D computational framework, WCSPH-ESPH for FSI problems, where structural fractures are modeled using the pseudo-spring analogy in ESPH, whereas the fluid is modeled using the weakly compressible SPH (WCSPH). Moreover, the interaction between the deformable structures and waters is modeled through modified generalized boundary conditions. In this approach, the interaction between the water and elastic structure particles is modeled by treating structure particles as dynamic boundaries for nearby fluid/ water particles. Fluid/ water pressure is extrapolated onto the structure particles to compute interaction forces, which are applied symmetrically based on Newton’s third law. The present contact model is similar to the no-slip boundary condition proposed by Adami et al. \cite{adami2012generalized} and does not require calibration of contact force like the soft repulsive particle contact model used in \cite{islam2024sph}. The proposed approaches have been rigorously validated through comparison with established computational methods and experimental data, affirming their accuracy and reliability. Furthermore, the system demonstrates a high level of proficiency in accurately modeling complex three-dimensional fracture patterns.

The rest of the paper is organized as follows: the WCSPH formulation for fluids is discussed in section 2. The ESPH formulation with pseudo-spring analogy is discussed in section 3. The detailed formulation of the calibration-free contact model is provided in section 4. Numerical results simulated with the present framework are discussed in section 5. We conclude our findings in section 6.

\section{Weakly-compressible SPH for fluids}
\subsection{Fundamental fomulations}
We solve the following evolution of the density and velocity fields for fluids over time, respectively, using SPH:
\begin{equation}
    \dfrac{d\rho}{dt} = -\rho \dfrac{\partial v^\beta}{\partial x^\beta},
\end{equation}

\begin{equation}
    \frac{d v^\alpha}{dt} = - \frac{1}{\rho} \frac{\partial p_r}{\partial x^\alpha} + \frac{1}{\rho} \frac{\partial \tau^{\alpha \beta}}{\partial x^\alpha} + g^\alpha,
\end{equation}

here, $\rho$ refers to the fluid density. In the Lagrangian frame of reference, $\displaystyle\frac{d}{d t}$ denotes the derivative with respect to time. The spatial coordinates at any given material point are denoted by $x^\alpha$, where $\alpha$ specifies the component index. Similarly, we represent the velocity component by $v^\alpha$. $g^\alpha$ represents the body force component in the $\alpha$ direction. The viscous stress is given by $\tau^{\alpha \beta}$, with $\alpha$ and $\beta$ implying the respective elements of the viscous stress:

\begin{equation}
    \tau^{\alpha \beta} = \mu_f \left( \frac{\partial v^\alpha}{\partial x^\beta} + \frac{\partial v^\beta}{\partial x^\alpha} \right),
\end{equation}
here, $\mu_f$ is the fluid viscosity. 

In this work, the pressure $p_r$ is derived using a weakly compressible equation of state, outlined as follows:

\begin{equation}
    p_r=p_0\left[\left(\frac{\rho}{\rho_0}\right)^\gamma-1\right]
\end{equation}
here, $\gamma$ is defined as a constant equal to 7 \cite{monaghan1994simulating}, $\rho_0$ denotes the reference density of the material, and $p_0$ is constant associated with fluctuations in fluid density. This specific constant is determined using the equation ${c^2_0 \rho_0}/{\gamma}$, with $c_0$ representing sound speed.

In SPH, we divide the domain into Lagrangian particles. Field variables for a particle at position $x^\alpha_i$ are computed using neighbouring particles at $x^\alpha_j$ with a kernel function $W(q,h)$, which approximates the Dirac-delta function. We denote the ratio of the normalized distance between a pair of particles (particles $i$ and $j$ with spatial coordinates $x^\alpha_i$ and $x^\alpha_j$ respectively) as $q=\norm{x^\alpha_i-x^\alpha_j} / h$, where $h$ is the smoothing length controlling the size of the support domain of the kernel function. This study utilizes the Wendland $C2$ kernel function \cite{wendland1995piecewise}. The equations governing the continuity and momentum can be written in its discrete SPH formulation as follows:

\begin{equation}
\label{eq:Mass}
\frac{d \rho_i}{d t}=\sum_j m_j v_{i j}^\beta W_{i j, \beta}+\delta h c_0 \sum_j 2 \frac{m_j}{\rho_j}\left(\rho_j-\rho_i + \rho_{ij}^H\right) \frac{x_{i j}^\beta}{\left\|x_i^\beta-x_j^\beta\right\|^2+0.01 h^2} W_{i j, \beta} \\
\end{equation}

\begin{equation}\label{acc_f}
\frac{d v_i^\alpha}{d t} = \sum_j m_j \left( \frac{\tau_i^{\alpha \beta}}{\rho_i^2} + \frac{\tau_j^{\alpha \beta}}{\rho_j^2} - \pi_{i j} \delta^{\alpha \beta} \right) W_{i j, \beta} - \sum_j m_j \left( \frac{p_{r_i}}{\rho_i^2} + \frac{p_{r_j}}{\rho_j^2} \right) W_{i j, \beta} + g^\alpha.
\end{equation} where $m_j$ is the mass of the $j$-th SPH particle, $v_{ij}^\beta = v_i^\beta - v_j^\beta$ is the difference in velocity, and $x_{ij}^\beta = x_i^\beta - x_j^\beta$ is the distance vector, $W_{ij,\beta}$ is the derivative of the kernel function in the $\beta$ spatial direction.

In the mass conservation equation (\ref{eq:Mass}), the first term on the right hand side (RHS) is the SPH discretization of the velocity divergence, and the second term is the $\delta$-SPH term employed to improve the pressure results \cite{fourtakas2019local}, in which $\delta$ is a constant taken as 0.1, and $\rho_{ij}^H$ is the density difference computed from the equation of state based on the hydrostatic pressure difference between the two particles \cite{fourtakas2019local}. This formulation is used to mitigate the instability at free surfaces in long-term simulations, which is caused by the truncated support domain.

In the momentum conservation equation (\ref{acc_f}), $\pi_{ij}$ represents the artificial viscosity used to prevent numerical instabilities and unphysical oscillations \cite{monaghan1983shock}. The following expression provides its definition:

\begin{equation}
\pi_{i j} = \begin{cases} 
\dfrac{-\beta_1 \bar{c}_{i j} \mu_{i j} + \beta_2 \mu_{i j}^2}{\bar{\rho}_{i j}}, & \text{if } v_{i j}^\alpha x_{i j}^\alpha \leq 0 \\ 
0, & \text{otherwise} 
\end{cases}
\end{equation}

where,

\begin{equation}
\mu_{i j} = \frac{h v_{i j}^\alpha x_{i j}^\alpha}{\left\|x_i^\alpha - x_j^\alpha\right\|^2 + 0.01 h^2}
\end{equation}
where $x^\alpha_{ij} =x^\alpha_i - x^\alpha_j$, $\bar{c}_{ij}$ denotes the average sound speed determined for particles $i$ and $j$, and $\bar{\rho}_{ij}=0.5 (\rho_i + \rho_j)$.

\subsection{No-slip boundary condition} \label{subsec:noslip}

\begin{figure}[htb!]
    \centering
\includegraphics[width=0.5\textwidth]{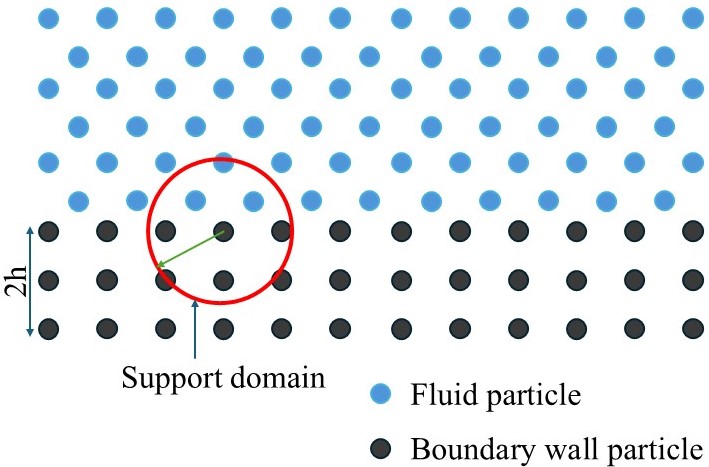}
    \caption{Diagram illustrating the boundary treatment applied at the solid wall, showing the interaction between the fluid and the solid surface.}
    \label{fig:boundary_treatment}
\end{figure}

The solid boundary is represented by boundary particles with extrapolated velocity and pressure information \cite{adami2012generalized}. These particles, represented in Figure \ref{fig:boundary_treatment}, have the same initial properties as fluid particles. The boundary particles participate in all SPH computations like regular particles; however, the field variables needed in the simulation are not computed using the governing equations. Instead, they are extrapolated from adjacent fluid particles. Among them, the computation of pressure needs to consider the influence of body forces:  
 
\begin{equation}\label{pr_ex}
    p_w=\dfrac{\sum_f p_f W(x_{wf})+(g^\beta-a_w^\beta) \sum_f \rho_f x_{wf}^\beta W(x_{wf})}{\sum_f W(x_{wf})}
\end{equation}
 
where the solid boundary wall particles are labeled with the subscript $w$, while fluid particles are represented by $f$. The variable $a_w^\beta$ signifies the $\beta$ component of the specified acceleration for the solid boundary wall particles. Consequently, the density of these wall particles is determined using the equation below:
\begin{equation}
    \rho_w=\rho_0\left(\frac{p_w}{p_0}+1\right)^{\frac{1}{\gamma}}
\end{equation}

\section{Pseudo-spring based SPH for solid deformation}

\subsection{SPH formulations for solid}
To model structural deformations, we solve the following governing equations: 
 
\begin{equation}
    \frac{d \rho}{d t}=-\rho \frac{\partial v^\beta}{\partial x^\beta},
\end{equation}
 
\begin{equation}
    \frac{d v^\alpha}{d t}=\frac{1}{\rho} \frac{\partial \sigma^{\alpha \beta}}{\partial x^\beta},
\end{equation}
 
where $\sigma^{\alpha \beta}$ is the Cauchy stress tensor associated with the indices $\alpha$ and $\beta$. Here, we solve the following discretized forms of the governing equations:
 
\begin{equation}
    \frac{d \rho_i}{d t}=\sum_j m_j v_{i j}^\beta W_{i j, \beta},
\end{equation}
 
\begin{equation}
    \frac{d v_i^\alpha}{d t}=\sum_j m_j\left(\frac{\sigma_i^{\alpha \beta}}{\rho_i^2}+\frac{\sigma_j^{\alpha \beta}}{\rho_j^2}-\pi_{i j} \delta^{\alpha \beta}-P_{i j}^a \delta^{\alpha \beta}\right) W_{i j, \beta}.
\end{equation}

The term $P_{ij}^a$ represents the artificial pressure correction, which aims to mitigate tensile instability \cite{monaghan2000sph}, which results in unphysical particle clumping and unrealistic cracks. To address this issue, the adjustment introduces a short-range repulsive force as described in the expression provided below:
  
\begin{equation}
    P_{i j}^a=\gamma\left(\frac{\left|p_{r_i}\right|}{\rho_i^2}+\frac{\left|p_{r_j}\right|}{\rho_j^2}\right)\left[\frac{W\left(d_{i j}\right)}{W(\Delta p)}\right]^{\bar{n}} \text {, }
\end{equation}
where $\gamma$ denotess the adjustment parameter, and $\bar{n}=W(0) / W(\Delta p)$, with $\Delta p$ representing the average particle spacing in the reference configuration.

\subsection{Constitutive model for elastic structure}
The stress tensor, denoted as $\sigma^{\alpha \beta}$, combines the effects of hydrostatic pressure, $p_r$, and deviatoric stress, $S^{\alpha \beta}$, where it is described by $\sigma^{\alpha \beta} = S^{\alpha \beta} - p_r \delta^{\alpha \beta}$. To compute the hydrostatic pressure in materials that can deform, a linear equation is used: $p_r = K (\rho/\rho_0 - 1)$ \cite{eliezer1986introduction}, with $K$ denoting the material's bulk modulus. The change in deviatoric stress $S^{\alpha \beta}$ with time is captured by the formula:
\begin{equation}
\label{dev_rate}
    \dot{S}^{\alpha \beta} = 2 \mu \left(\dot{\epsilon}^{\alpha \beta} - \frac{1}{3} \delta^{\alpha \beta} \dot{e}^{\gamma \gamma}\right) + S^{\alpha \gamma} \omega^{\beta \gamma} + S^{\gamma \beta} \omega^{\alpha \gamma},
\end{equation}
where $\mu$ represents the shear modulus. The first term on the RHS is the equation from the linear elasticity theory, whereas the second and third terms are from the Jaumann stress rate, which is used to maintain the material objectivity in large deformation and rotation \cite{gray2001sph}. The formulations for computing the rate of strain tensor $\dot{\epsilon}^{\alpha \beta}$ and the spin tensor $\omega^{\alpha \beta}$ are
\begin{align}
\label{strain_rate}
    \dot{\epsilon}^{\alpha \beta} &= \frac{1}{2} \left(l^{\alpha \beta} + l^{\beta \alpha}\right), \\
    \omega^{\alpha \beta} &= \frac{1}{2} \left(l^{\alpha \beta} - l^{\beta \alpha}\right),
\end{align}
where the velocity gradient \(l^{\alpha \beta}\) is derived as:
\begin{equation}
    l^{\alpha \beta} = \sum_j \left(v_j^{\alpha} - v_i^{\alpha}\right) W_{i j,\beta} \frac{m_j}{\rho_j}.
\end{equation}

\subsection{Damage and failure modeling}
The kernel functions in SPH peak near the centre of their support domain and diminish rapidly with increasing distance. Consequently, particles close to a reference particle $i$ significantly influence the interaction compared to those at the kernel's outer edges \cite{chakraborty2013pseudo, islam2017computational}. In our study, for the solid modeling, we estimate the field variables at particle $i$ by summing contributions only from its nearest neighbours, i.e., we only consider particles that can connect to $i$ directly without intersecting others. This treatment is necessary because the pseudo-spring method considers only connections between intermediate neighbours \cite{chakraborty2013pseudo,islam2017computational,islam2019total}.

Since only the intermediate neighbours are considered in the solid simulation, the conventional SPH approximations of field functions and their gradients are inaccurate. To mitigate this problem, we apply a gradient correction method to reduce truncation errors due to incomplete support domains. We replace $W_{ij,\beta}$ with $\hat{W}_{ij,\beta}$, which is calculated as: 
\begin{equation}
    \hat{W}_{i j, \beta} = B_i^{\beta a} W_{i j, a} \quad \text{with} \quad \mathbf{B}_{i} = \mathbf{A}_{i}^{-1} \quad \text{and} \quad A_i^{\beta a} = -\sum_j \frac{m_j}{\rho_j} x_{i j}^\beta W_{i j, a}.
\end{equation}

\begin{figure}[htb!]
    \centering
    \begin{subfigure}[t]{0.48\textwidth}
        \centering
        \includegraphics[width=\textwidth]{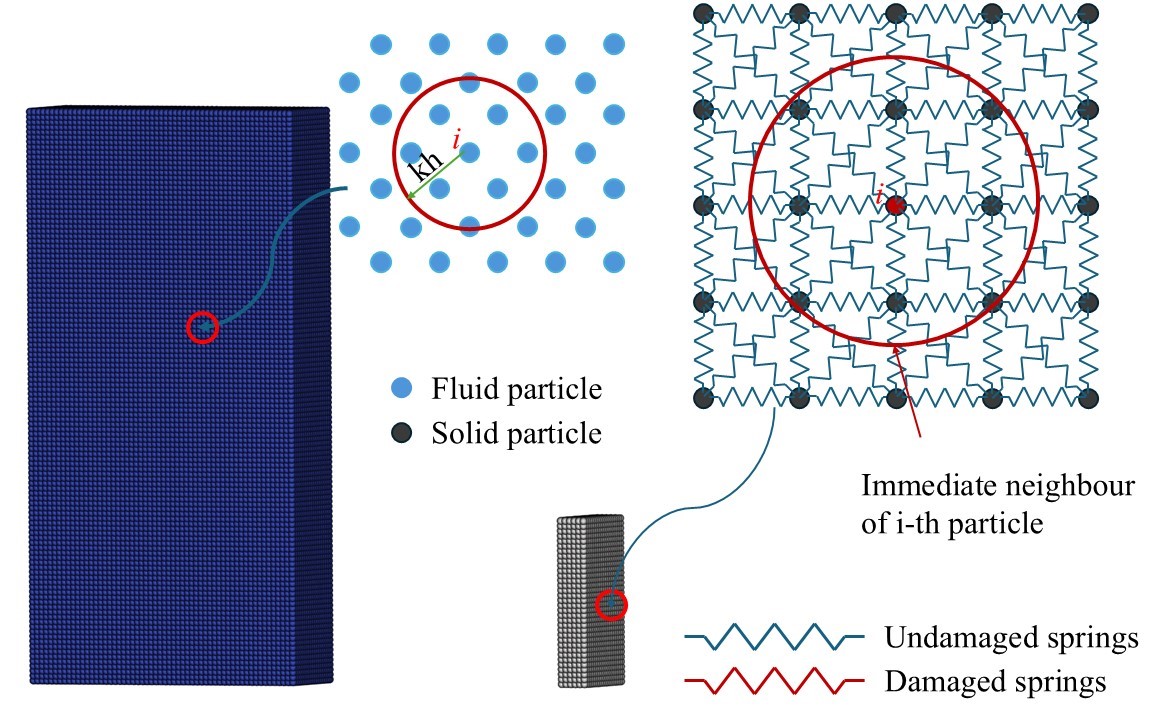}
        \caption{Undamaged pseudo-spring network before fluid impact.}
        \label{image1}
    \end{subfigure}
    \hfill
    \begin{subfigure}[t]{0.48\textwidth}
        \centering
        \includegraphics[width=\textwidth]{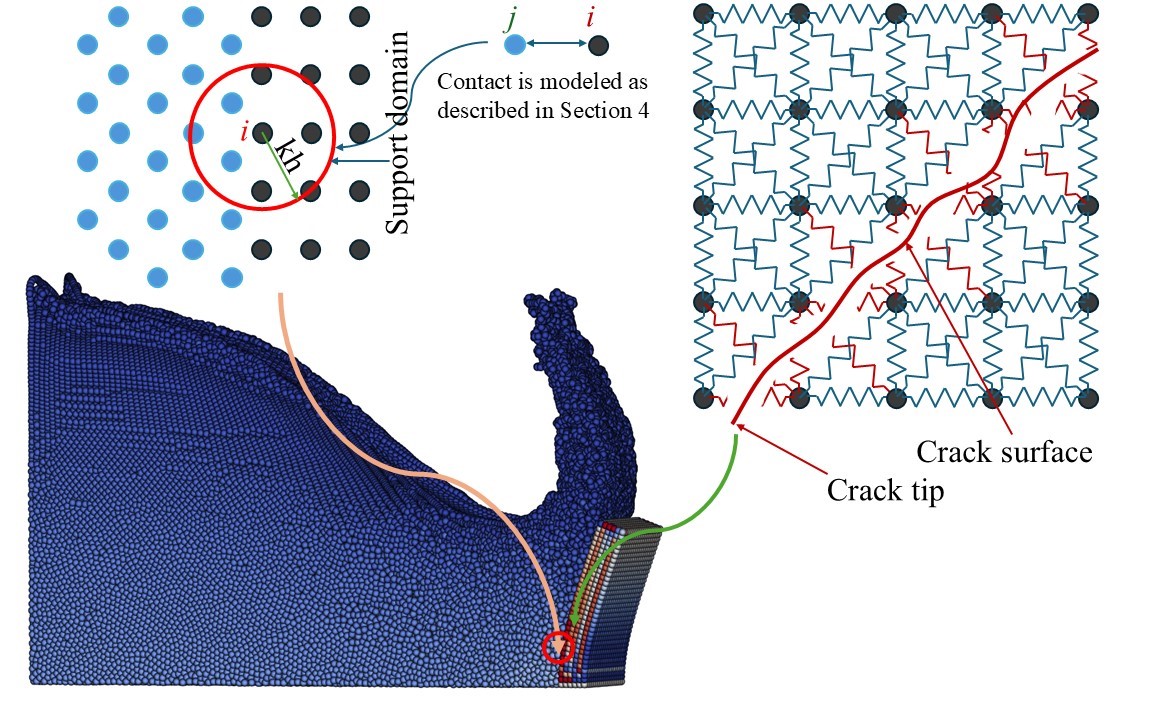}
        \caption{Fractured pseudo-springs indicating crack formation due to fluid-solid interaction.}
        \label{image2}
    \end{subfigure}
    \caption{Modeling crack evolution using the pseudo-spring analogy in SPH. (a) Initial state with intact connections; (b) Crack propagation visualized through broken pseudo-springs following fluid impact.}
    \label{fig:pseudo_crack}
\end{figure}

The nearest neighbouring particles are connected to the $i$-th particle via pseudo springs as shown in Fig. \ref{fig:pseudo_crack}. These springs do not add stiffness but define interaction levels $f_{ij}$. If the material is undamaged, $f_{ij}=1$; upon reaching critical stress or strain, it permanently becomes $f_{ij}=0$. We compute the damage index in the pseudo-springs as $D_{ij} = 1 - f_{ij}$, i.e., permanently damaged pseudo-springs are indicated by $D_{ij} = 0$. The formation of permanently damaged or failed pseudo-springs provides a reliable means to track the evolution of the crack path within the computational domain. To reflect evolving interactions due to these pseudo springs, we modify the kernel functions and their gradients to incorporate the interaction level $f_{ij}$ by using $f_{ij} \hat{W}_{ij}$ and $f_{ij} \hat{W}_{ij, \beta}$. The modified conservation equations are:
 
\begin{equation}
    \frac{d \rho_i}{d t} = \sum_{j \in \mathbb{N}_U^i} m_j v_{ij}^\beta \hat{W}_{ij, \beta} + \sum_{j \in \mathbb{N}_D^i} m_j v_{ij}^\beta (f_{ij} \hat{W}_{ij, \beta}),
\end{equation}

\begin{equation}\label{acc_s}
\begin{aligned}
& \frac{d v_i^a}{d t} = \sum_{j \in \mathbb{N}_U^i} m_j \left(\frac{\sigma_i^{a \beta}}{\rho_i^2} + \frac{\sigma_j^{a \beta}}{\rho_j^2} - \pi_{ij} \delta^{a \beta} - P_{ij}^a \delta^{a \beta}\right) \hat{W}_{ij, \beta} \\
& + \sum_{j \in \mathbb{N}_D^i} m_j \left(\frac{\sigma_i^{a \beta}}{\rho_i^2} + \frac{\sigma_j^{a \beta}}{\rho_j^2} - \pi_{ij} \delta^{\alpha \beta} - P_{ij}^a \delta^{a \beta}\right) (f_{ij} \hat{W}_{ij, \beta}),
\end{aligned}
\end{equation}
where $\mathbb{N}_D^i$ includes damaged neighbors and $\mathbb{N}_U^i$ includes undamaged ones. To visualize crack propagation, we use a damage variable $D$ defined as the ratio of failed pseudo springs to total springs connected to a particle. $D = 1$ indicates complete failure, while values between $0$ and $1$ represent partial damage. Cracks can propagate even when $D_i < 1$, indicating that damage can affect connections beyond individual particles. For details of the pseudo-spring method, interested readers can refer to \cite{chakraborty2013pseudo,islam2017computational}.

\section{Coupling between fluid flow and deformable structures}\label{interaction}
This section outlines the approach for linking the two solvers: WCSPH (used for fluid simulation) and pseudo-spring SPH (used for elastic structures with possible brittle cracks). The solid and fluid phases are discretized with uniform initial particle spacing.

In the simulations, the particles of the elastic structure (indexed by $a$) serve as boundary particles for the fluid particles (indexed by $i$) near the fluid-structure interface. Following the boundary treatment method discussed before, we extrapolate the fluid pressure ($p_r$) at each structure particle from neighbouring fluid particles using Eq. \ref{pr_ex}. Thereafter, we update the boundary particles' densities using the equation of state. At this stage, it is important to emphasize that this pressure and density are applied solely for computing interactions within the fluid particles and do not represent the physical properties of the elastic structure. Consequently, we can write the following equation to calculate the force $f_i^s$ exerted on the $i$-th fluid particle by the neighbouring structure particles as

\begin{equation}
    f_i^s = -m_i \sum_a m_a \left( \frac{p_i}{\rho_i^2} + \frac{p_a}{\rho_a^2} - \pi_{ia} \delta_{ia} \right) W_{ia,\beta}
\end{equation}

Following Newton’s third law, a force $f_a^F$ of equal magnitude but opposite direction will be exerted on each structure particle involved in the two-phase interaction. We can calculate $f_a^F$ exerted on deformable solid particles as

\begin{equation}
    f_a^F = -m_a \sum_i m_a \left( \frac{p_a}{\rho_a^2} + \frac{p_i}{\rho_i^2} - \pi_{ai} \delta_{ai} \right) W_{ia,\beta}.
\end{equation}

Finally, we calculate the acceleration of each particle involved in the two-phase interaction and update Eqs. (\ref{acc_f}) and (\ref{acc_s}) respectively to model the two-phase interaction accurately. A predictor-corrector integration method is used for temporal integration as described in Algorithm~\ref{alg:time_integration}. The solid and fluid solvers utilize the same time step based on the Courant-Friedrichs-Lewy (CFL) condition to ensure consistency and avoid discrepancies in time integration.

\begin{algorithm}
\caption{ESPH-WCSPH: One Computational Step Using Predictor–Corrector Time Integration Scheme}
\label{alg:time_integration}
\begin{algorithmic}[1]
\State \textbf{Input:} Velocity and position of fluid and structure particles at time step $n$: $\mathbf{v}_i^n$, $\mathbf{v}_a^n$, $\mathbf{x}_i^n$, $\mathbf{x}_a^n$.
\State Update neighbour lists for fluid and structure particles based on the current configuration
\State Extrapolate pressure at structure particles $p_a$ from neighboring fluid particles and compute density $\rho_a$ using the equation of state
\State Compute velocity rate $\dot{\mathbf{v}}_i^n$ using Eq.~\eqref{acc_f}, and density rate $\dot{\rho}_i^n$ using Eq.~\eqref{eq:Mass} for fluid particles
\State Compute strain rate $\dot{\epsilon}^{\alpha \beta}$ using Eq.~\eqref{strain_rate} and deviatoric stress tensor rate $\dot{S}^{\alpha \beta}$ using Eq.~\eqref{dev_rate} for structure particles

\State \textbf{Predictor Step:}
\[
\begin{aligned}
\mathbf{v}_i^{n+1/2} &= \mathbf{v}_i^n + \frac{1}{2} \Delta t\, \mathbf{f}_i^n, \quad 
\mathbf{x}_i^{n+1/2} = \mathbf{x}_i^n + \frac{1}{2} \Delta t\, \mathbf{v}_i^n \\
\mathbf{v}_a^{n+1/2} &= \mathbf{v}_a^n + \frac{1}{2} \Delta t\, \mathbf{f}_a^n, \quad 
\mathbf{x}_a^{n+1/2} = \mathbf{x}_a^n + \frac{1}{2} \Delta t\, \mathbf{v}_a^n \\
\boldsymbol{\epsilon}_a^{n+1/2} &= \boldsymbol{\epsilon}_a^n + \frac{1}{2} \Delta t\, \dot{\boldsymbol{\epsilon}}_a^n, \quad 
\mathbf{S}_a^{n+1/2} = \mathbf{S}_a^n + \frac{1}{2} \Delta t\, \dot{\mathbf{S}}_a^n
\end{aligned}
\]

\State \textbf{Predict new density and pressure for fluid particles:}
\[
\rho_i^{n+1/2} = \rho_i^n + \frac{1}{2} \Delta t\, \dot{\rho}_i^n, \quad 
p_i^{n+1/2} = p_0 \left[ \left( \frac{\rho_i^{n+1/2}}{\rho_0} \right)^\gamma - 1 \right]
\]

\State Repeat steps 3 to 8 with predicted values to evaluate mid-step forces and rates

\State \textbf{Corrector Step:}
\[
\begin{aligned}
\mathbf{v}_i^{n+1/2} &= \mathbf{v}_i^n + \frac{1}{2} \Delta t\, \mathbf{f}_i^{n+1/2}, \quad 
\mathbf{x}_i^{n+1/2} = \mathbf{x}_i^n + \frac{1}{2} \Delta t\, \mathbf{v}_i^{n+1/2} \\
\mathbf{v}_a^{n+1/2} &= \mathbf{v}_a^n + \frac{1}{2} \Delta t\, \mathbf{f}_a^{n+1/2}, \quad 
\mathbf{x}_a^{n+1/2} = \mathbf{x}_a^n + \frac{1}{2} \Delta t\, \mathbf{v}_a^{n+1/2} \\
\boldsymbol{\epsilon}_a^{n+1/2} &= \boldsymbol{\epsilon}_a^n + \frac{1}{2} \Delta t\, \dot{\boldsymbol{\epsilon}}_a^{n+1/2}, \quad 
\mathbf{S}_a^{n+1/2} = \mathbf{S}_a^n + \frac{1}{2} \Delta t\, \dot{\mathbf{S}}_a^{n+1/2}
\end{aligned}
\]

\State \textbf{Correct density and pressure:}
\[
\rho_i^{n+1/2} = \rho_i^n + \Delta t\, \dot{\rho}_i^{n+1/2}, \quad 
p_i^{n+1/2} = p_0 \left[ \left( \frac{\rho_i^{n+1/2}}{\rho_0} \right)^\gamma - 1 \right]
\]

\State \textbf{Final Update:} $t \gets t + \Delta t$
\[
\begin{aligned}
\mathbf{v}_i^{n+1} &= 2\mathbf{v}_i^{n+1/2} - \mathbf{v}_i^n, \quad 
\mathbf{x}_i^{n+1} = 2\mathbf{x}_i^{n+1/2} - \mathbf{x}_i^n, \quad 
\rho_i^{n+1} = 2\rho_i^{n+1/2} - \rho_i^n \\
\mathbf{v}_a^{n+1} &= 2\mathbf{v}_a^{n+1/2} - \mathbf{v}_a^n, \quad 
\mathbf{x}_a^{n+1} = 2\mathbf{x}_a^{n+1/2} - \mathbf{x}_a^n \\
\boldsymbol{\epsilon}_a^{n+1} &= 2\boldsymbol{\epsilon}_a^{n+1/2} - \boldsymbol{\epsilon}_a^n, \quad 
\mathbf{S}_a^{n+1} = 2\mathbf{S}_a^{n+1/2} - \mathbf{S}_a^n
\end{aligned}
\]
\end{algorithmic}
\end{algorithm}

\section{Numerical examples}

In this section, several numerical examples are presented to validate the method from different aspects. The solver is accelerated using GPU parallelisation based on LOQUAT \cite{peng2019loquat}. All simulations are performed using a desktop equipped with RTX 4090 graphics card. In the current work, we will not investigate the numerical efficiency, as it is studied in some related papers \cite{zhan2019stabilized,zhan2019three}.

\subsection{3D dam break flow}
\begin{figure}[hbtp!] 
\centering
\includegraphics[width=0.8\textwidth]{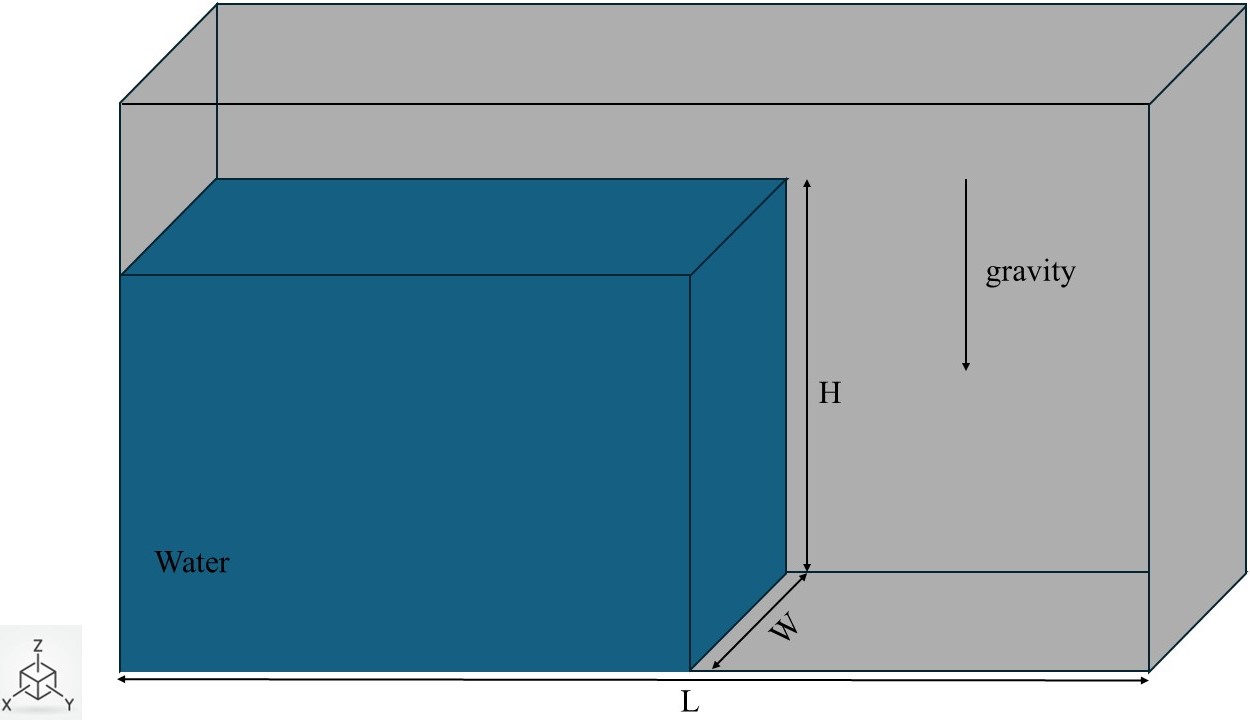}
\caption{Geometry and boundary conditions for the dam break test setup} 
\label{water_setup}
\end{figure}

\begin{figure}[hbtp!]
\centering

\begin{subfigure}[t]{\textwidth}
    \includegraphics[width=\textwidth]{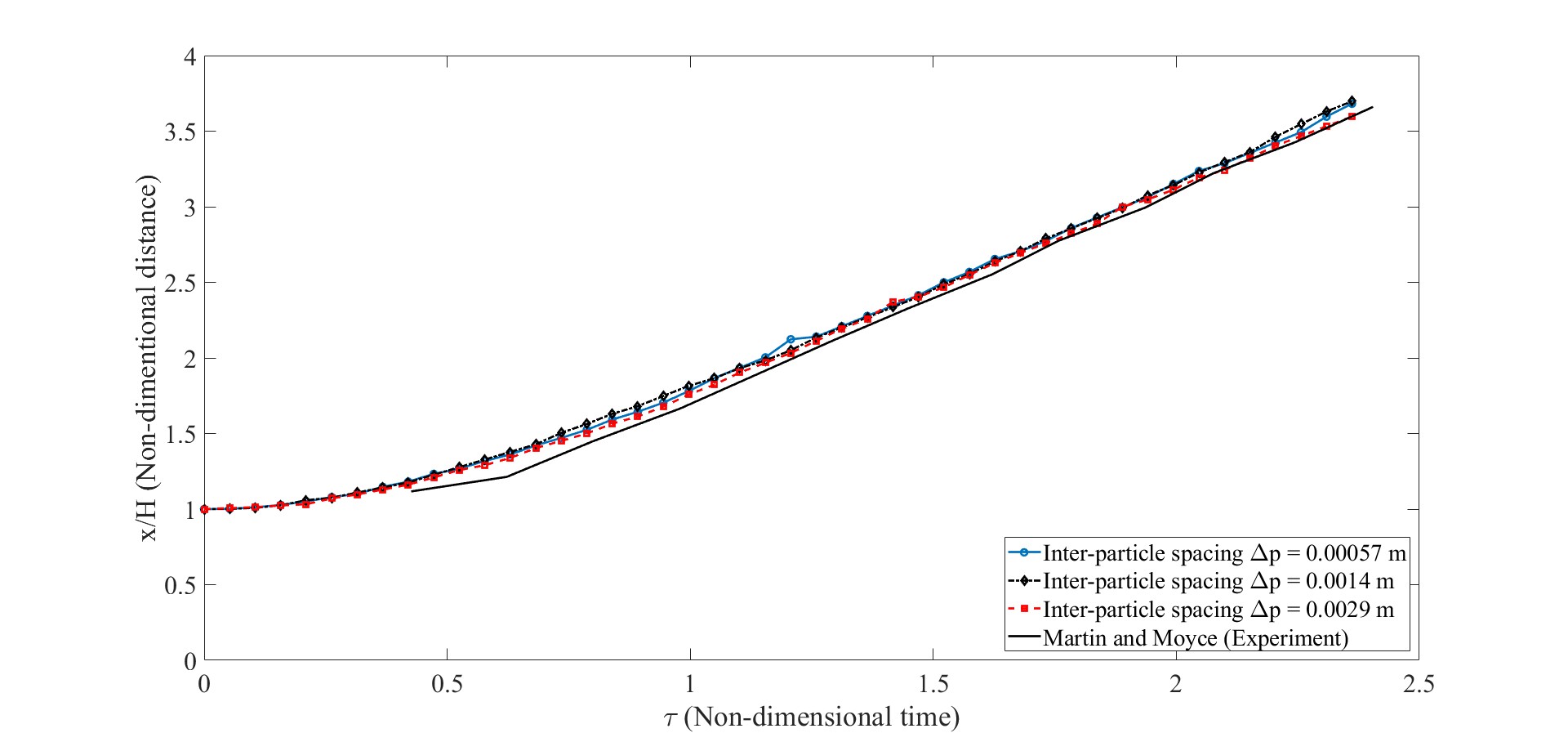}
    \caption{Effect of varying inter-particle spacing $\Delta p$}
    \label{fig:dam_dp}
\end{subfigure}

\vspace{1em} 

\begin{subfigure}[t]{\textwidth}
    \includegraphics[width=\textwidth]{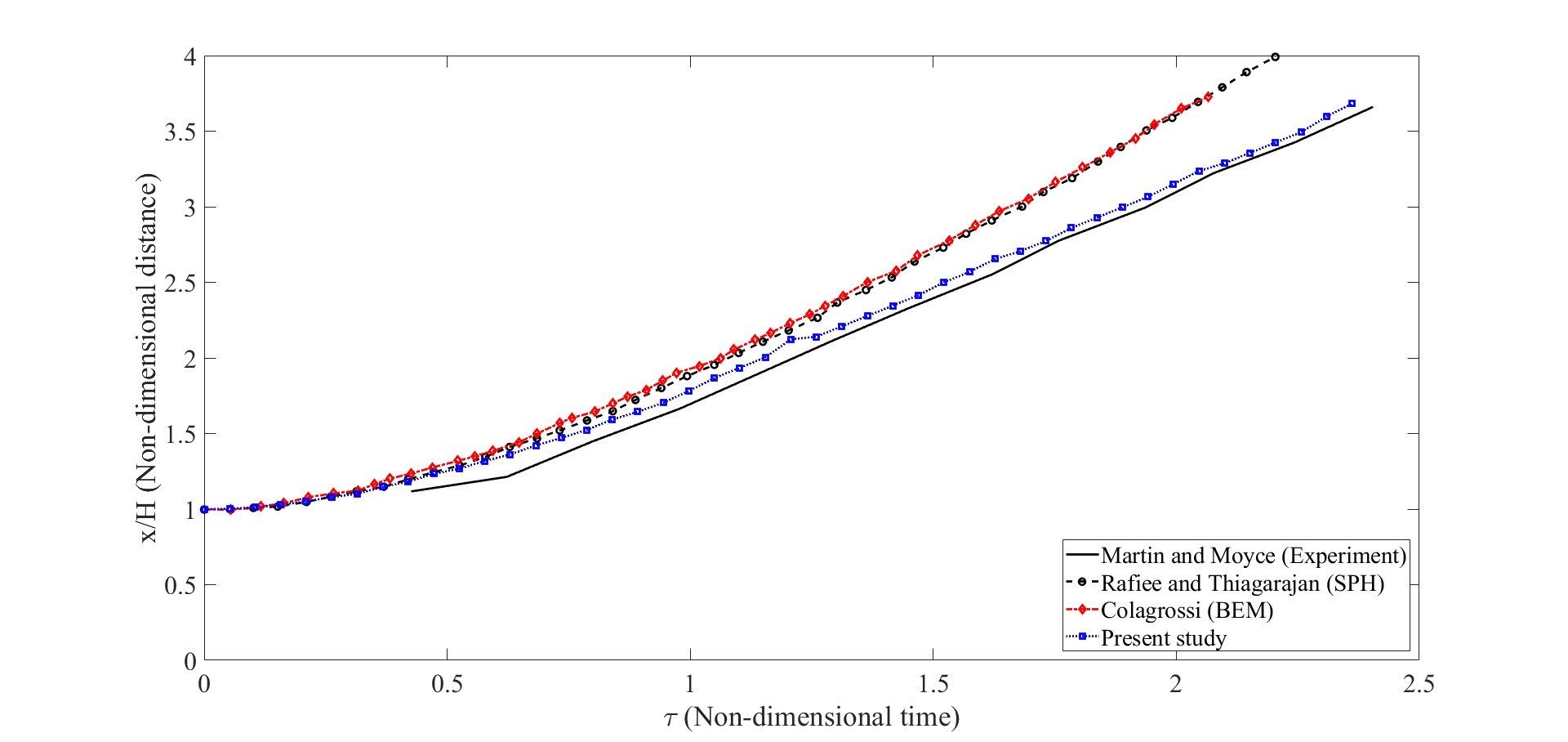}
    \caption{Comparison with other results at $\Delta p = 0.00057\,\text{m}$
}
    \label{fig:dam_comp}
\end{subfigure}

\caption{Time history of the water-front toe location, referenced from the left boundary wall, and the impact of varying inter-particle distances on its evolution in the dam break simulation ($\tau = \frac{t}{\sqrt(H/g)}$)}
\label{fig:dam_vertical}
\end{figure}

\begin{figure}[hbtp!] 
\centering
\begin{subfigure}[t]{0.49\textwidth}
\includegraphics[width=\textwidth,trim={325 125 50 150}, clip]{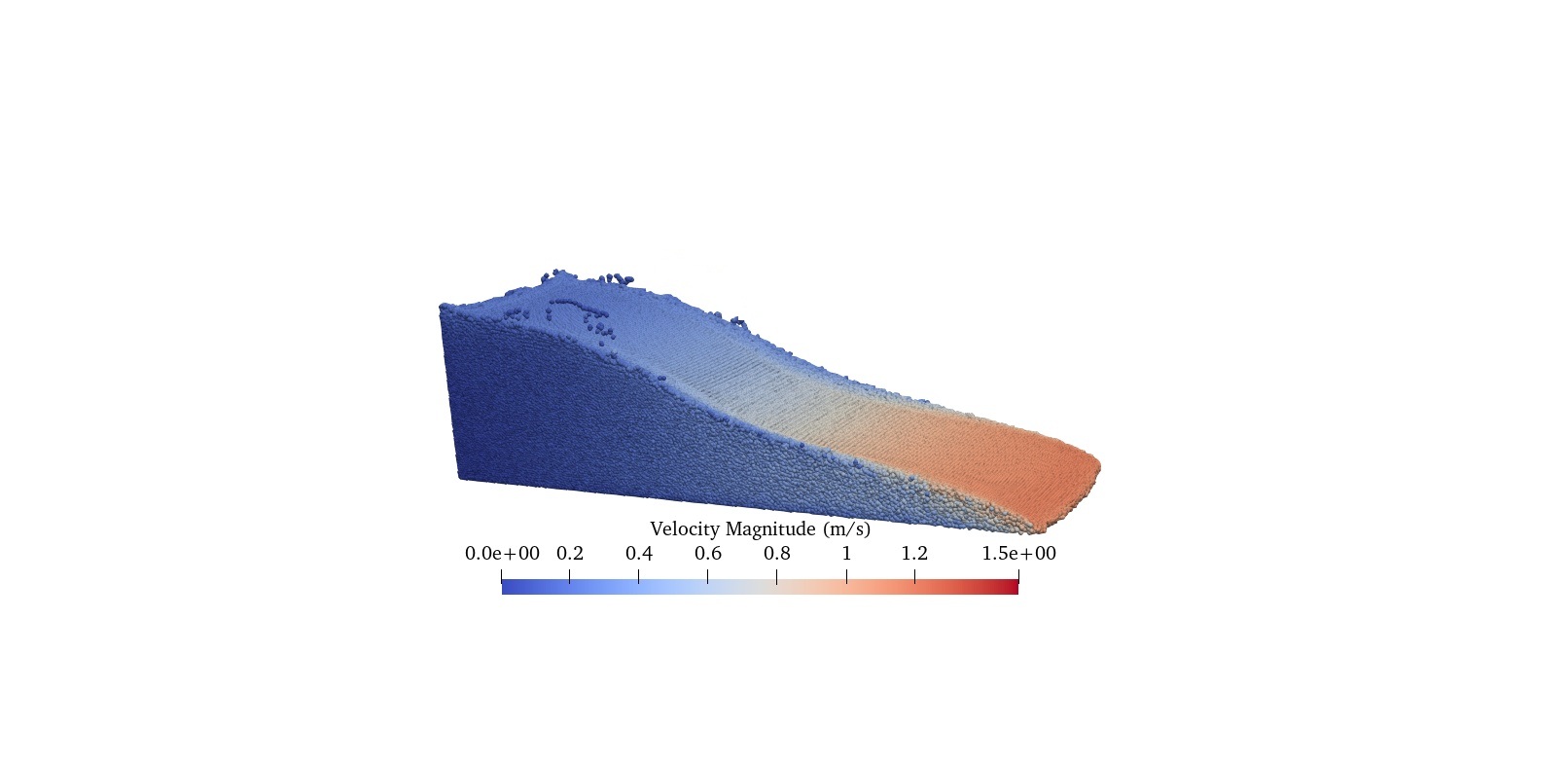}
\caption{Time = 0.12 s} 
\end{subfigure}
\begin{subfigure}[t]{0.49\textwidth}
\includegraphics[width=\textwidth,trim={325 125 50 150}, clip]{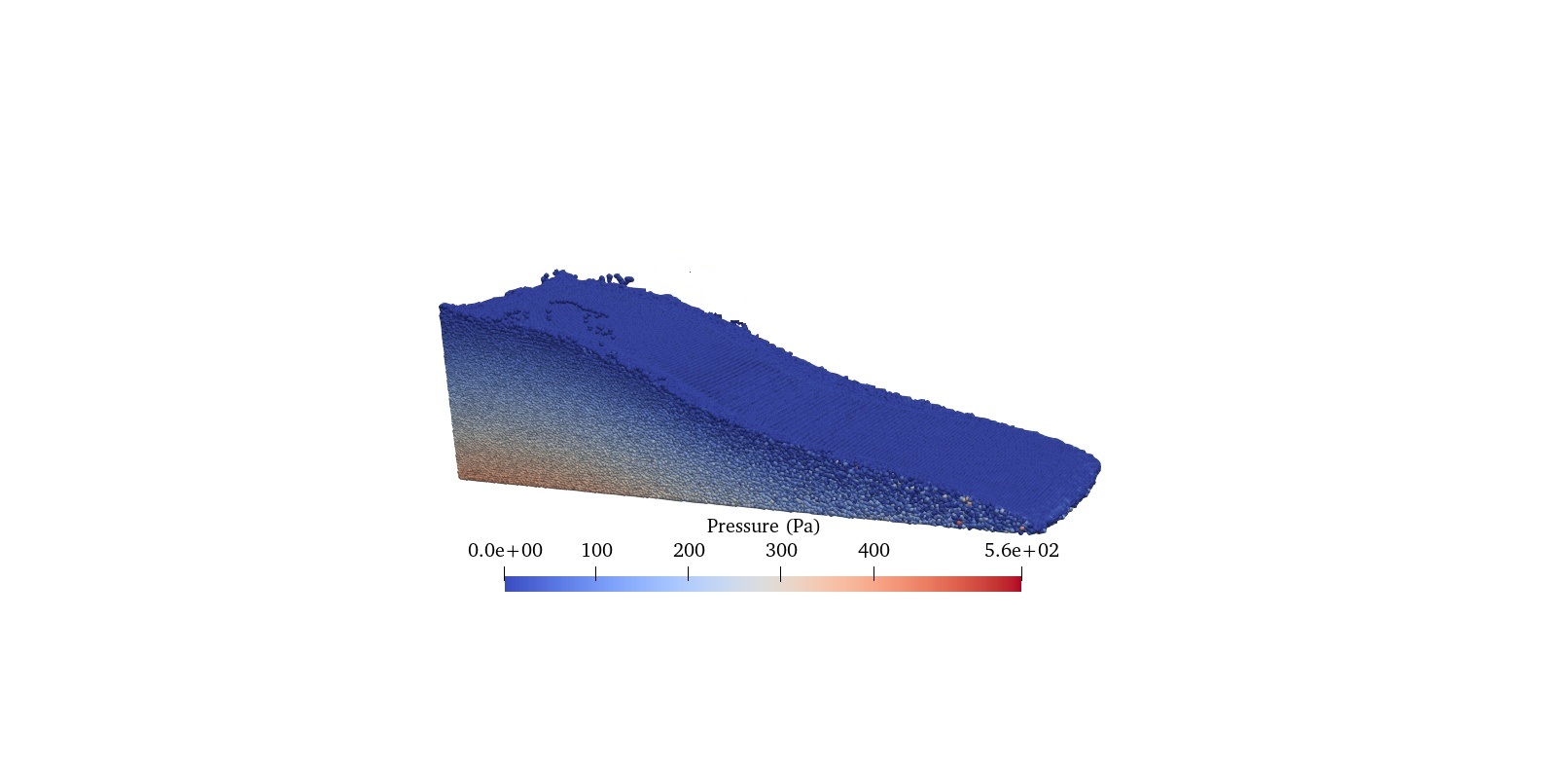}
\caption{Time T=0.12 s} 
\end{subfigure}
\begin{subfigure}[t]{0.49\textwidth}
\includegraphics[width=\textwidth,trim={325 125 50 150}, clip]{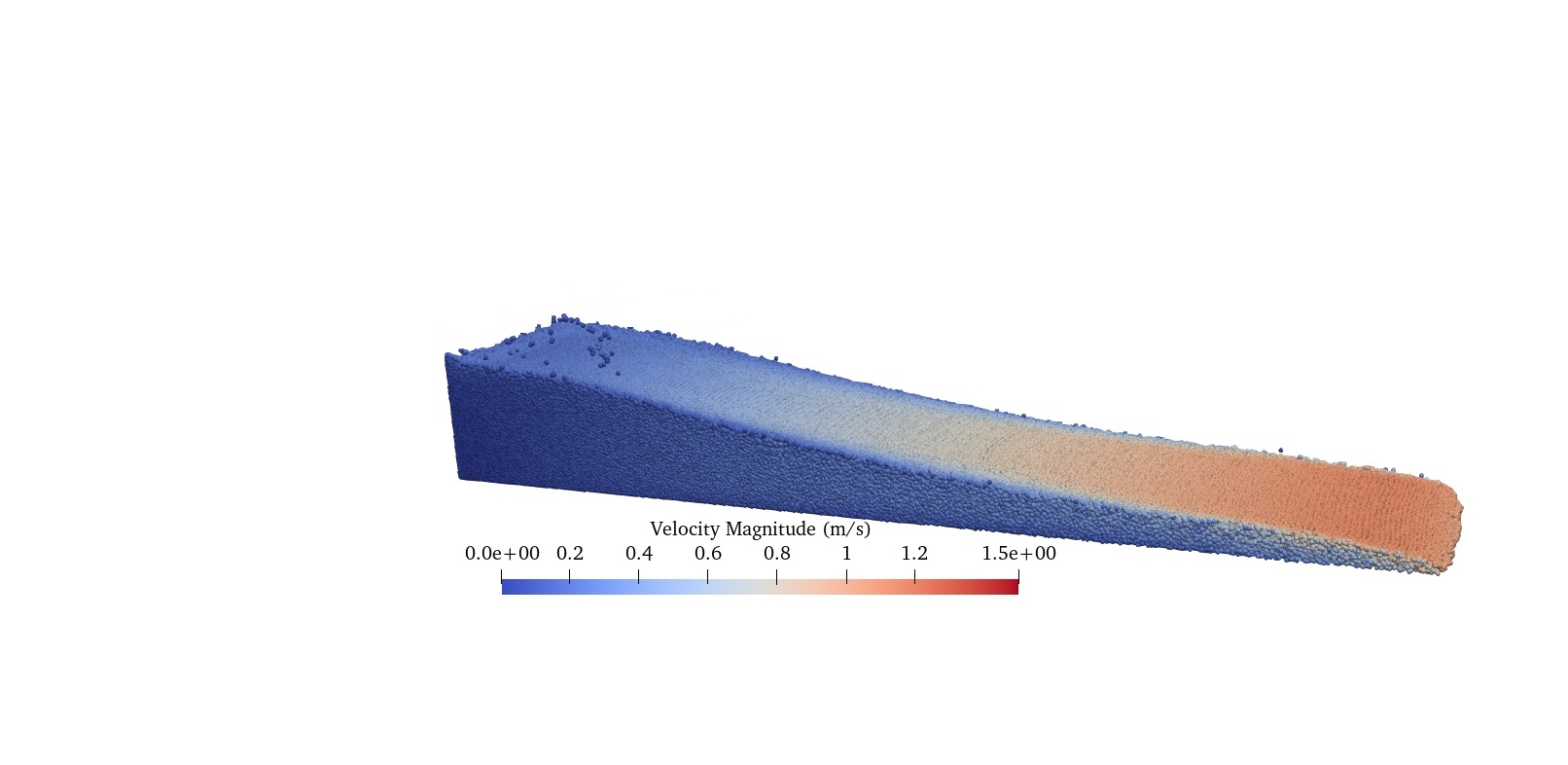}
\caption{Time T=0.19 s} 
\end{subfigure}
\begin{subfigure}[t]{0.49\textwidth}
\includegraphics[width=\textwidth,trim={325 125 50 150}, clip]{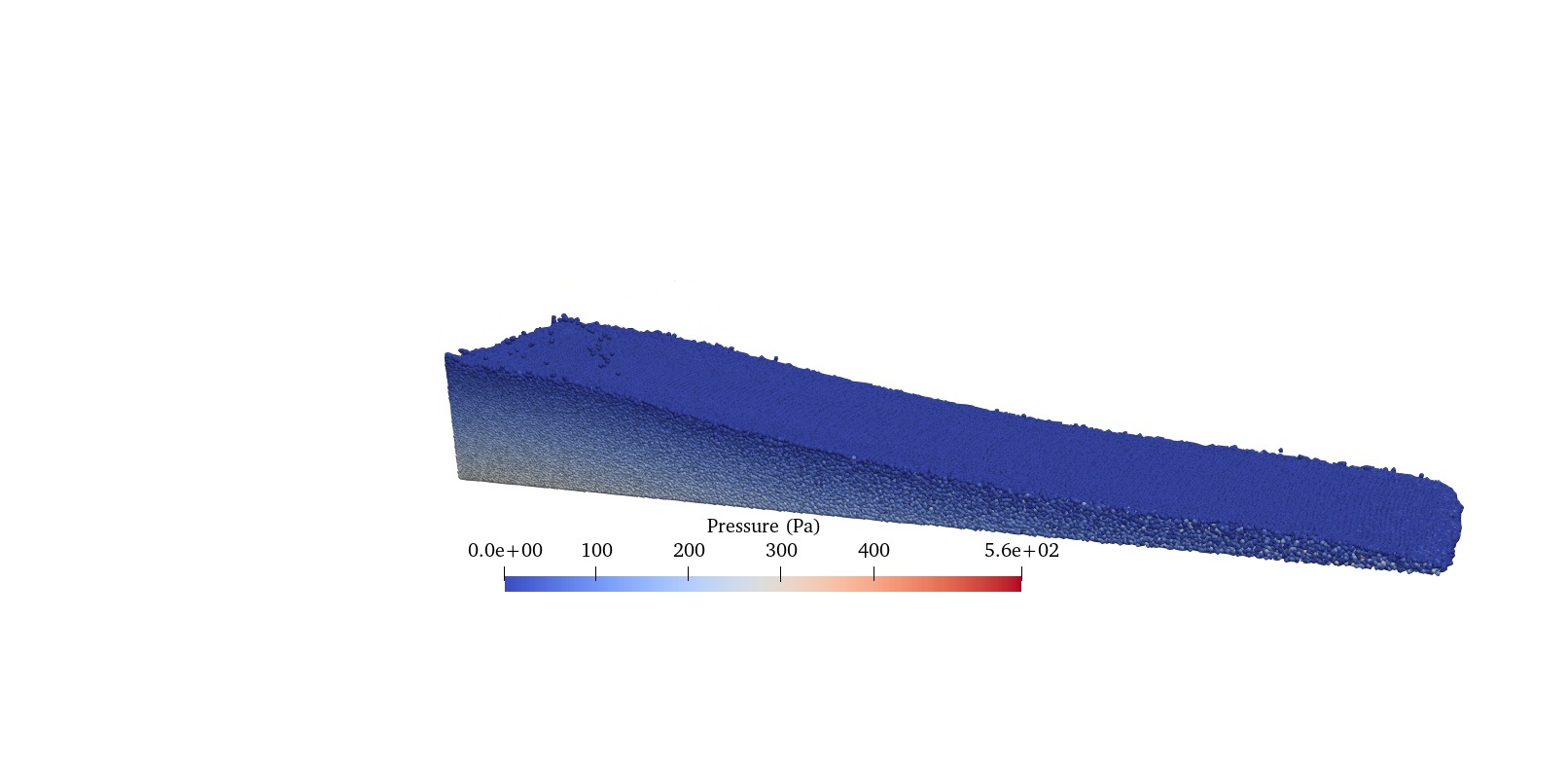}
\caption{Time T=0.19 s} 
\end{subfigure}
\begin{subfigure}[t]{0.49\textwidth}
\includegraphics[width=\textwidth,trim={325 125 50 150}, clip]{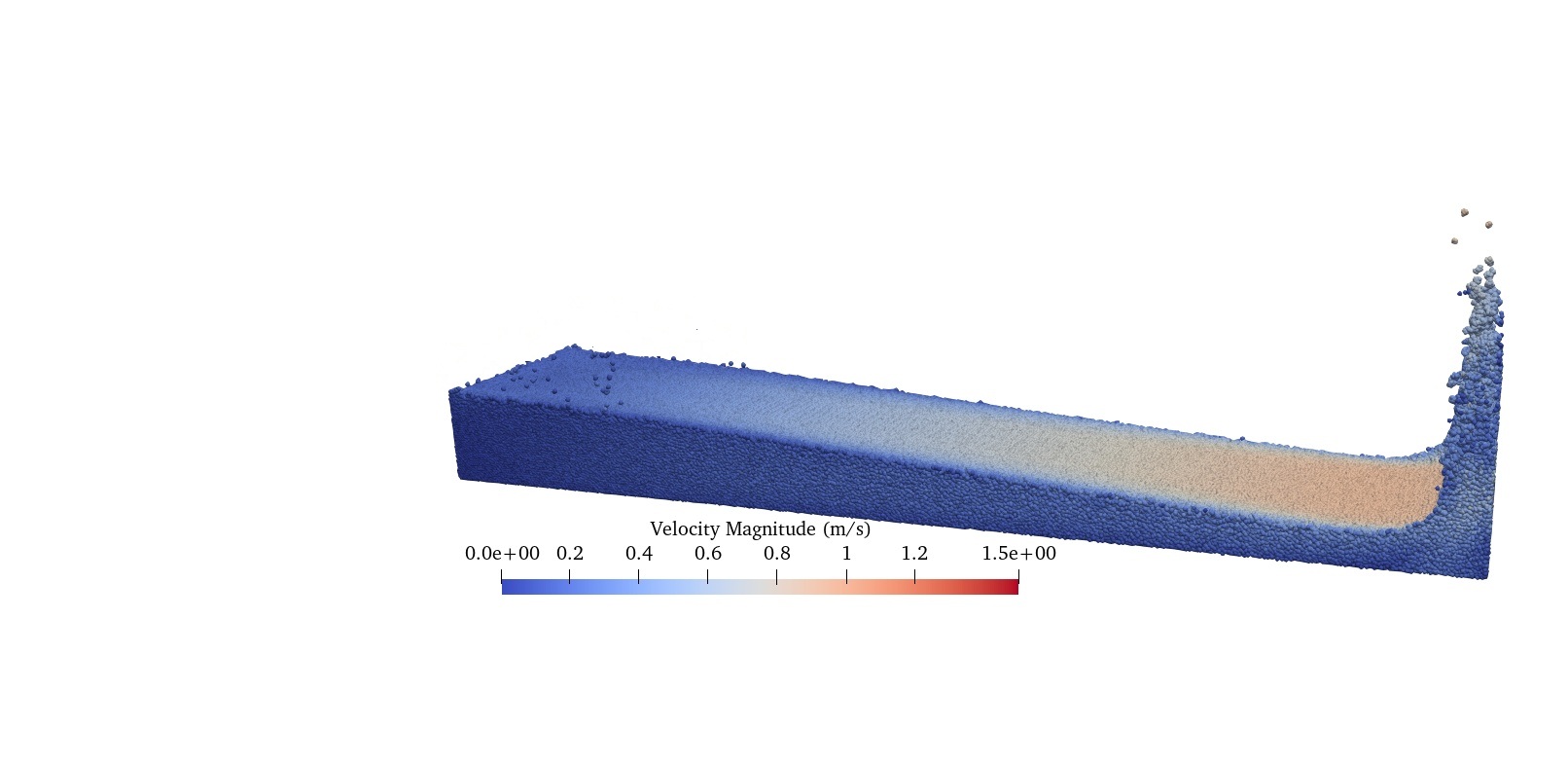}
\caption{Time T=0.25 s} 
\end{subfigure}
\begin{subfigure}[t]{0.49\textwidth}
\includegraphics[width=\textwidth,trim={325 125 50 150}, clip]{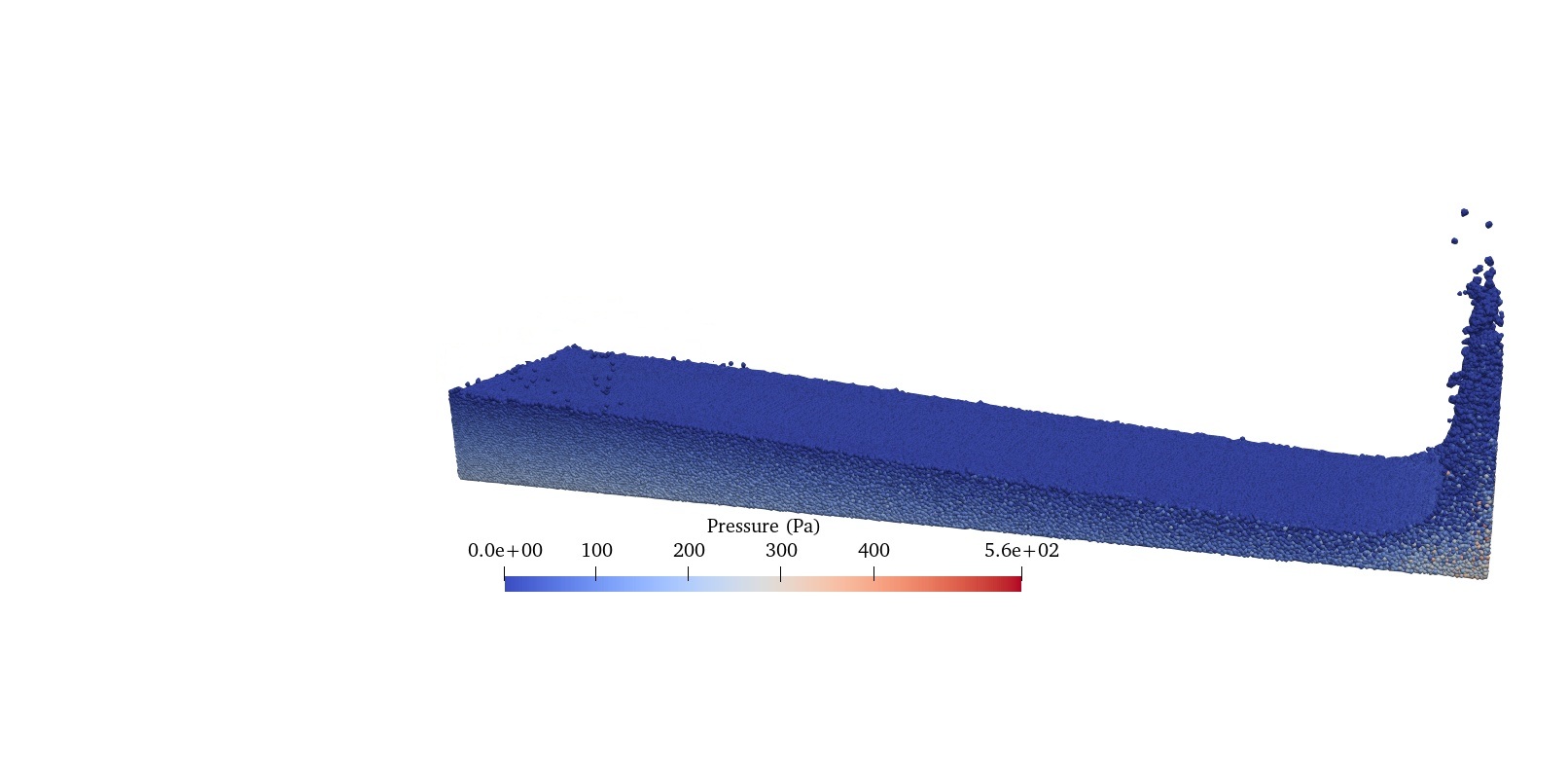}
\caption{Time T=0.25 s} 
\end{subfigure}
\caption{Velocity magnitude and pressure distribution countours at 0.12\,s, 0.19\,s, and 0.25\,s in the dam break test}
\label{velocitycountour}
\end{figure}

In this example, we tested our developed code against a benchmark case of 3D dam break flow. Our results are compared with the experimental data \cite{martin1952part} and established numerical studies \cite{colagrossi2005meshless, rafiee2009sph} which are available in the literature. The test configuration, depicted in Fig. \ref{water_setup}, consists of an initial rectangular domain with $W=H=0.057~m$ and $L=4H$. The water is modeled with a density of 1000 $kg/m^3$ and a dynamic viscosity of $\mu_f=0.05~Pa-s$. The boundary condition specified in section \ref{subsec:noslip} governs the interaction between the rigid wall and the fluid. To study the effect of inter-particle spacing, $\Delta p$, simulations are performed using three inter-particle spacings: $\Delta p = 0.00057~m$, $0.0014~m$ and $0.0029~m$. The non-dimensional time ($\tau = \frac{t}{\sqrt(H/g)}$) vs non-dimensional distance ($\frac{x}{H}$, $x$ being the position of the most forward water particles) with $\frac{h}{\Delta p} =1.3$ and $\Delta t=1~\mu s$ is illustrated in Fig. \ref{fig:dam_vertical} and also compared with the experimental results \cite{martin1952part}. We can see that the results for different $\Delta p$ are comparable with the experimental result. We have also compared the non-dimensional distance ($\frac{x}{H}$) with the other results available in the literature and can readily observe that they agree well. Fig. \ref{velocitycountour} shows the velocity and pressure contours at selected time steps. The simulation captures key features of free-surface flow during dam break, including water propagation over a dry bed, impact with a vertical wall, and subsequent rise and overturning. The results are consistent with previous studies \cite{martin1952part, colagrossi2005meshless, rafiee2009sph}.

\subsection{3D crack branching under mode-I loading}

\begin{figure}[htb!]
    \centering
    \includegraphics[width=\textwidth,trim={50 50 40 50}, clip]{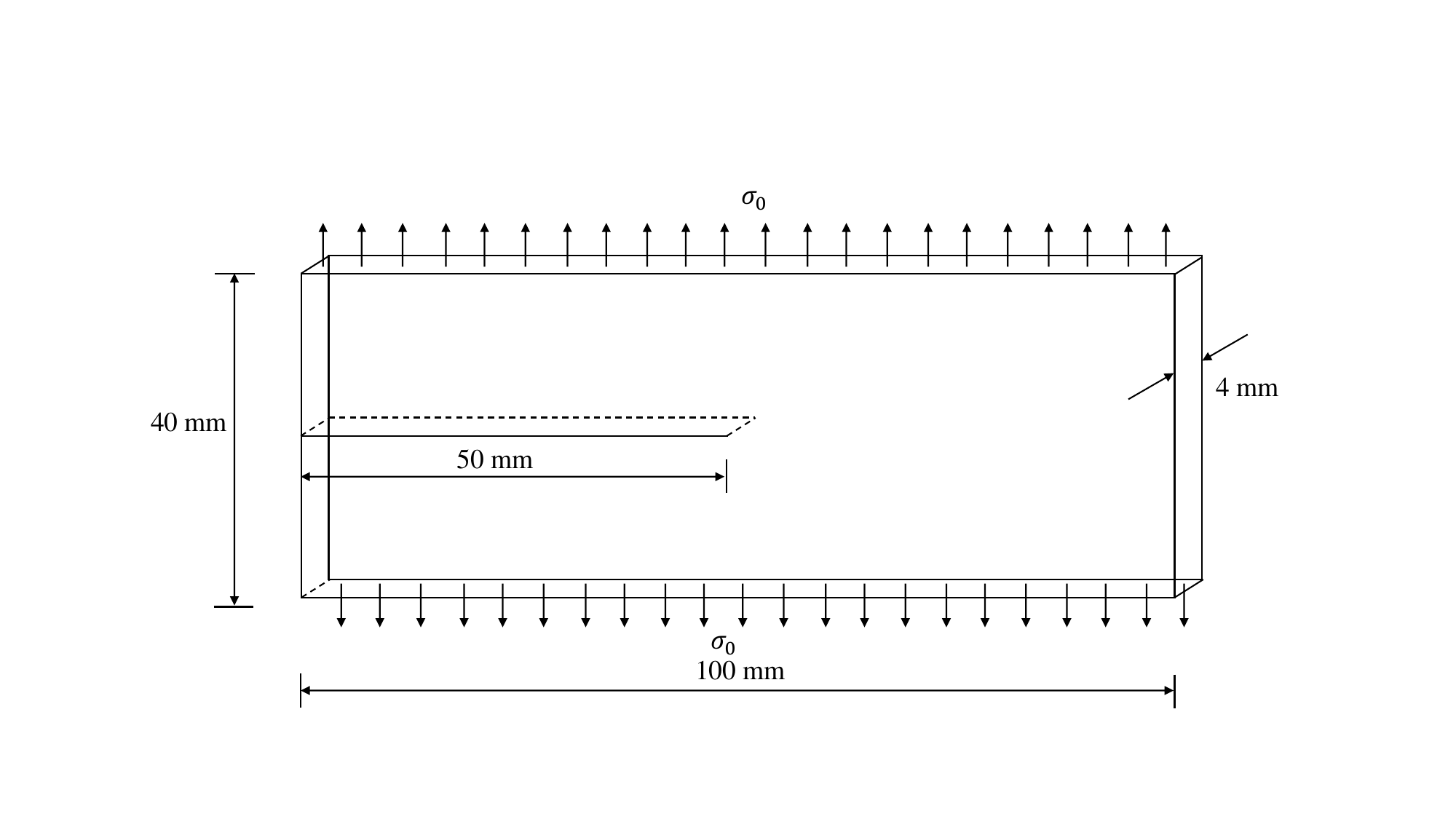}
       \caption{Setup of pre-notched brittle plate under tensile loading}
       \label{crack_branching}
\end{figure}

\begin{figure}[hbtp!] 
\centering
\begin{subfigure}[t]{\textwidth}
\includegraphics[width=\textwidth,trim={20 15 20 15}, clip]{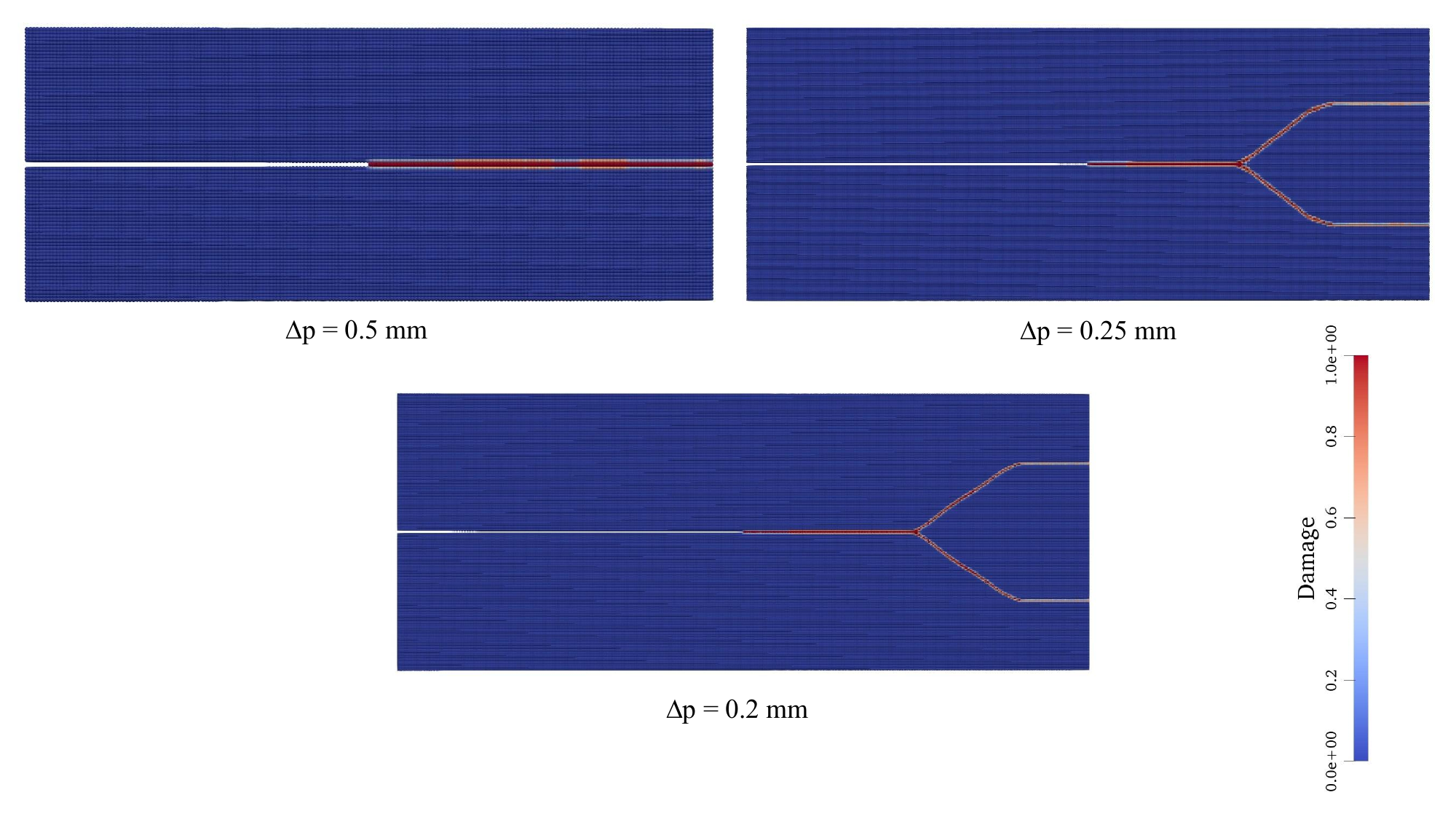}
\caption{} 
\end{subfigure}
\caption{Comparison of the crack path under tensile loading ($\sigma$ = 1 MPa): }
\label{2d_plane}
\end{figure}

In order to compare our results with those of Refs. \cite{song2009dynamic, agwai2011predicting, song2008influence}, we use their values of material parameters ($E=32~GPa$, $\rho_0 = 2450~kg/m^3$, $\nu=0.2$, and failure strain $\epsilon_{max} = 0.000509$), and study 3D deformations of a 100 mm x 40 mm x 4 mm plate with a 50 mm side-notch located at its centre as shown in Fig. \ref{crack_branching}. We apply 1 MPa tensile stress on the top and bottom surfaces of the plate for the crack to propagate under mode-I loading. The crack propagation speed, $v_{crack}$ is computed from

\begin{equation}
    v_{crack} = \frac{|| \bm{x}_t - \bm{x}_{t-\Delta t}||}{\Delta t}
\end{equation}
where $\bm{x}_t$ is the crack-tip location at time t. For different inter-particle spacing, $\Delta p$ = 0.5, 0.25 and 0.2 mm, the crack initiation and branching times are listed in Table \ref{table_branching_compa_dp_3d}. We also compare the final crack paths for different $\Delta p$ in Fig. \ref{2d_plane}. For particle spacing $\Delta p$ = 0.25 mm and 0.2 mm, the crack initiates at 14 $\mu s$ and 12 $\mu s$ respectively, and the crack branching happens at 52 $\mu s$ for both cases. The crack branching does not occur for $\Delta p$ = 0.5 mm. For the remaining simulations, we use $\Delta p$ = 0.2 mm. For $\frac{h}{\Delta p}$ = 1.3, 1.5, and 1.7, the crack initiates at 12 $\mu$s and the crack branching occurs at $\approx$ 52 $\mu$s for all the cases. We use $\frac{h}{\Delta p}$ = 1.5 henceforth. We have also performed a convergence study with different $\Delta t$ = 0.01, 0.005 and 0.001 $\mu$s (with $\frac{h}{\Delta p}$ = 1.5, $\Delta p$ = 0.2 mm). For the three values of $\Delta t$, the crack initiates at 12 $\mu$s and branches at about 52$\mu s$.

\begin{table}[h!]
\caption{Comparison of times at which crack initiation and branching occur for different $\Delta p$ in 3D}\label{table_branching_compa_dp_3d}
\centering
\begin{tabular}{cccc}
\hline
  & $\Delta p$ = 0.5 mm & $\Delta p$ = 0.25 mm  & $\Delta p$ = 0.2 mm \\ \hline
Crack initiation & & &\\
 time ($\mu$s) in 3D & 22 & 14 & 12   \\ \hline
Crack branching & & & \\          
 time ($\mu$s) in 3D & No branching & 52 & 52 \\ \hline
\end{tabular}
\end{table}

\begin{figure}[hbtp!] 
\centering
\begin{subfigure}[t]{0.3\textwidth}
\includegraphics[width=\textwidth,trim={20 15 20 15}, clip]{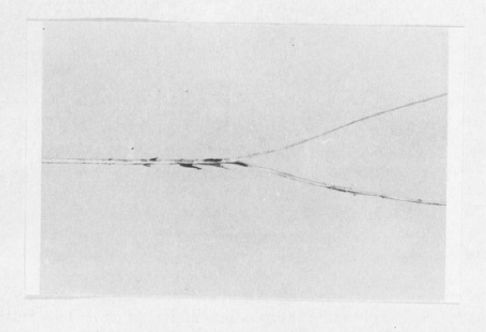}
\caption{}\label{exp_br}
\end{subfigure}
\begin{subfigure}[t]{0.49\textwidth}
\includegraphics[width=\textwidth]{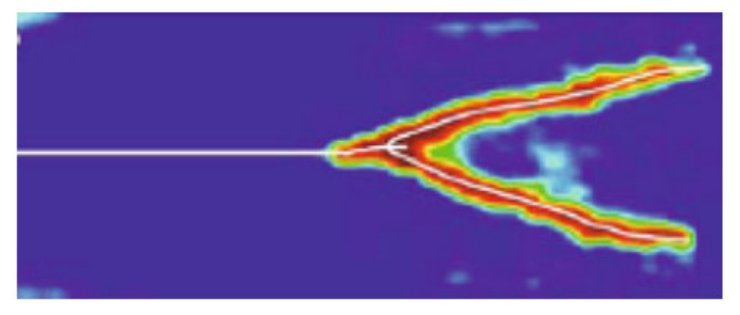}
\caption{} 
\end{subfigure}
\begin{subfigure}[t]{0.49\textwidth}
\includegraphics[width=\textwidth,trim={20 100 20 50}, clip]{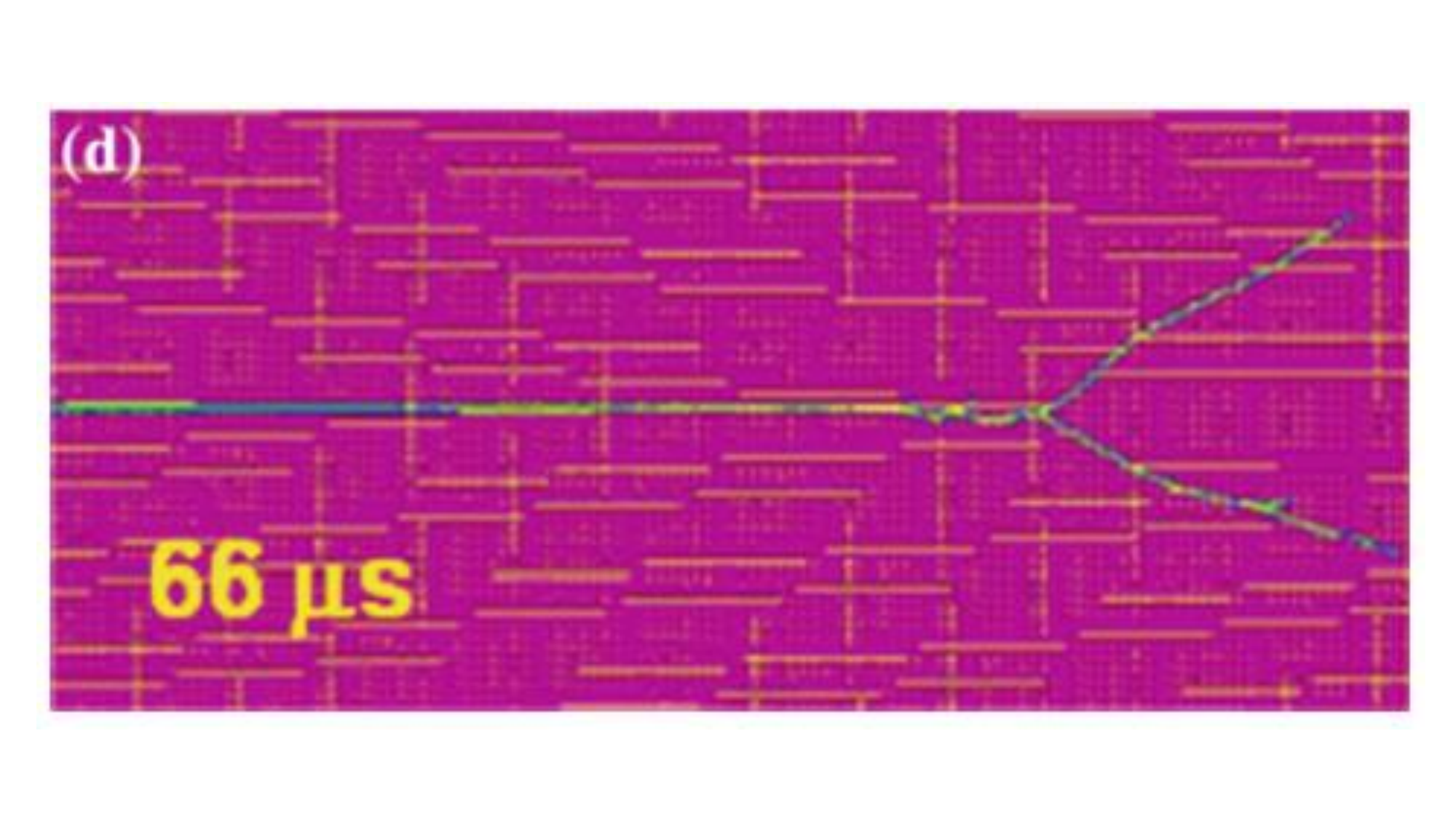}
\caption{} 
\end{subfigure}
\begin{subfigure}[t]{0.49\textwidth}
\includegraphics[width=\textwidth,trim={10 10 10 10}, clip]{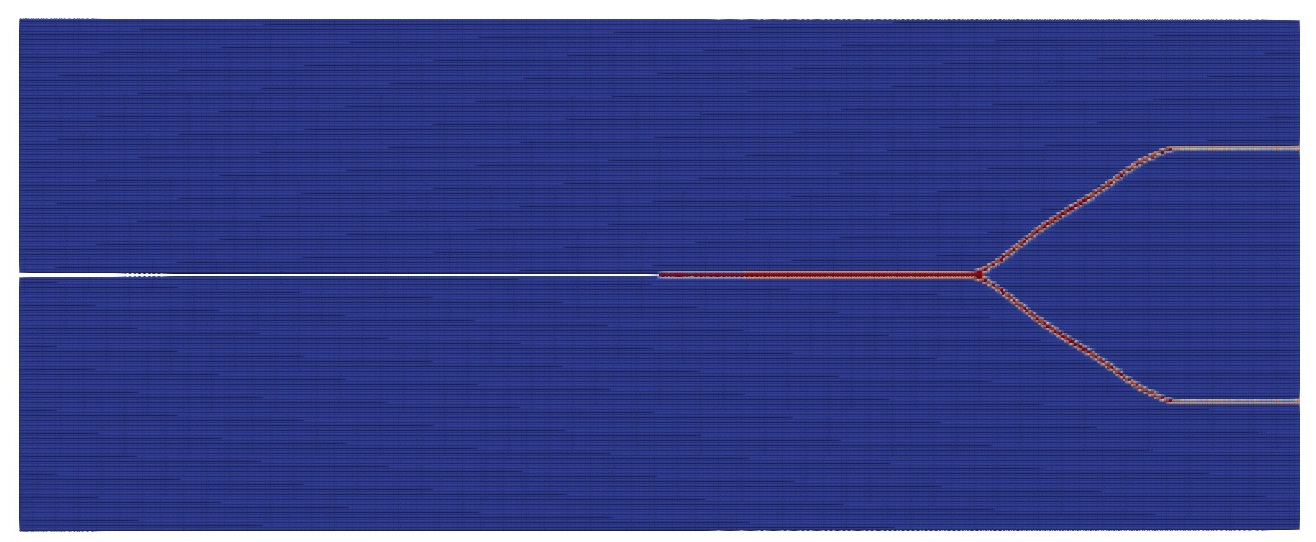}
\caption{} 
\end{subfigure}
\begin{subfigure}[t]{0.7\textwidth}
\includegraphics[width=\textwidth,trim={10 10 10 10}, clip]{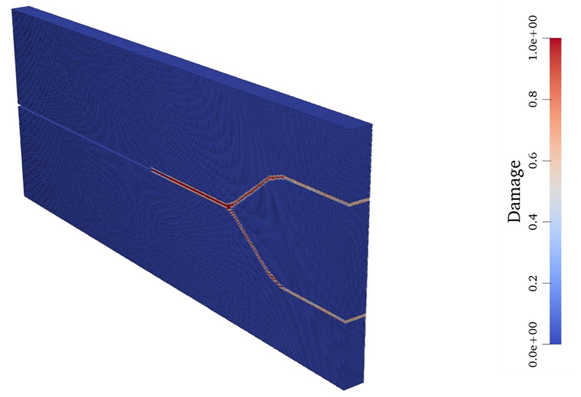}
\caption{}\label{br_3d} 
\end{subfigure}
\caption{Comparison of the crack path under tensile loading ($\sigma$ = 1 MPa): (a) Experiment \cite{ramulu1985mechanics}, (b) XFEM \cite{song2008comparative}, (c) Peridynamics \cite{agwai2011predicting} (d) Present and (e) Present 3D crack path}\label{branching_path_comp}
\end{figure}

\begin{figure}[htb!]
    \centering
    \includegraphics[width=\textwidth]{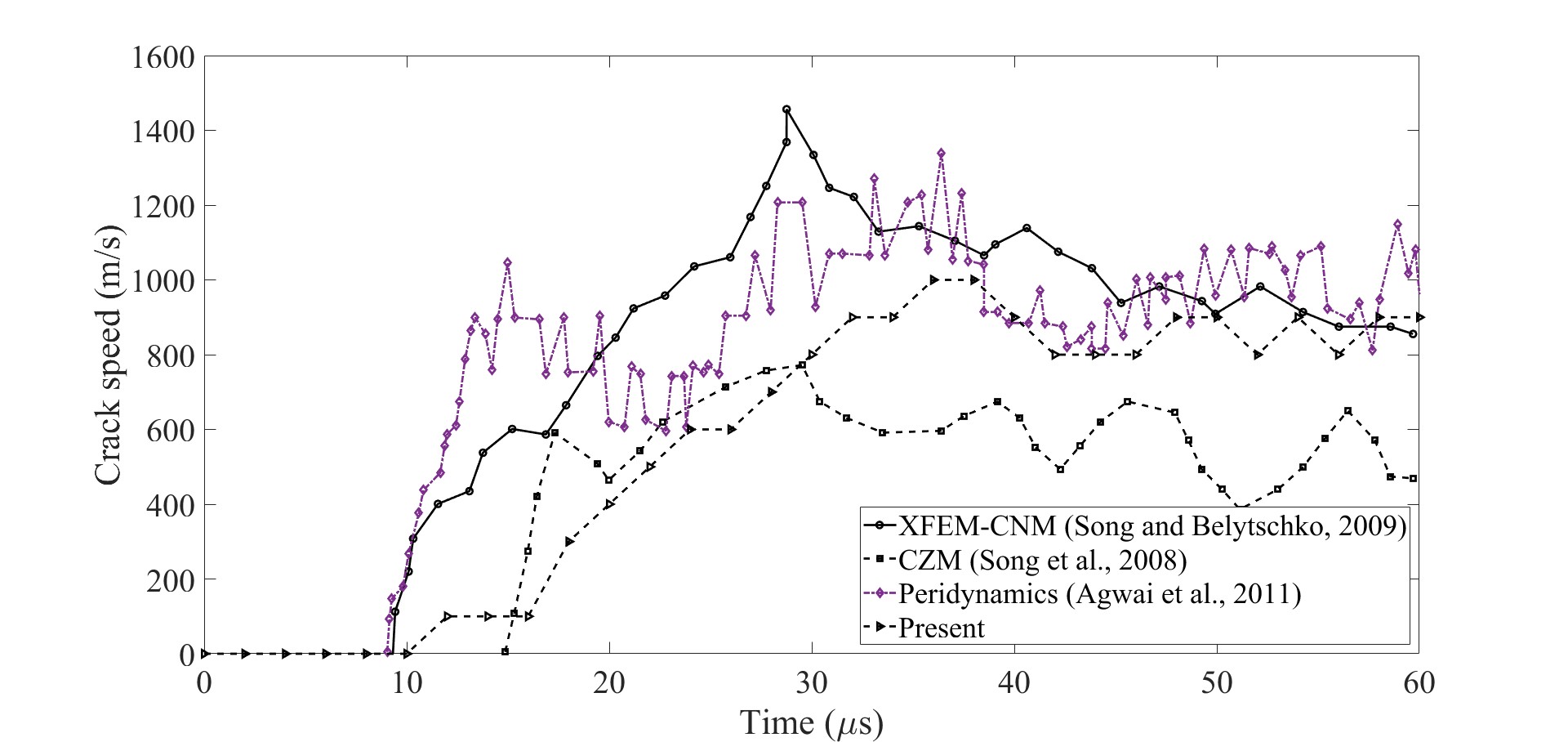}
       \caption{Comparison of crack propagation speed over time for mode I tensile loading (Fig.~\ref{crack_branching}).}
       \label{Crack_speed}
\end{figure}

We have also compared our results with experimental and numerical results available in the literature. 
In the experiment \cite{ramulu1985mechanics}, the crack initiates from the notch-tip, propagated straight for some time and then split into two parts that propagated to the edge of the plate (Fig. \ref{exp_br}). We have compared the crack path obtained from the present framework with other numerical results: XFEM \cite{song2008comparative} and Peridynamics \cite{agwai2011predicting} in Fig. \ref{branching_path_comp}. The presently computed crack path replicates major characteristics observed experimentally and numerically. The computed crack path, exhibited in Fig. \ref{branching_path_comp}, is qualitatively similar to that observed in experiments. The 3D crack path from our simulation is shown in Fig/ \ref{br_3d}. The time history of the crack speed is compared with that of other investigators in Fig. \ref{Crack_speed} and they are found to be comparable.

\subsection{Dam breaking flow through a rubber gate}
A dam-break case studied in \cite{antoci2007numerical, rafiee2009sph, zhan2019stabilized, sun2020smoothed, salehizadeh2022coupled} is simulated, where the structure failure is not considered. With this numerical example, we aim to validate the structure and fluid solvers and the coupling approach. Particularly, the accuracy of the structure solver with only intermediate neighbors and corrected kernel gradient is checked.

\begin{figure}[htb!]
    \centering
    \includegraphics[width=0.7\textwidth]{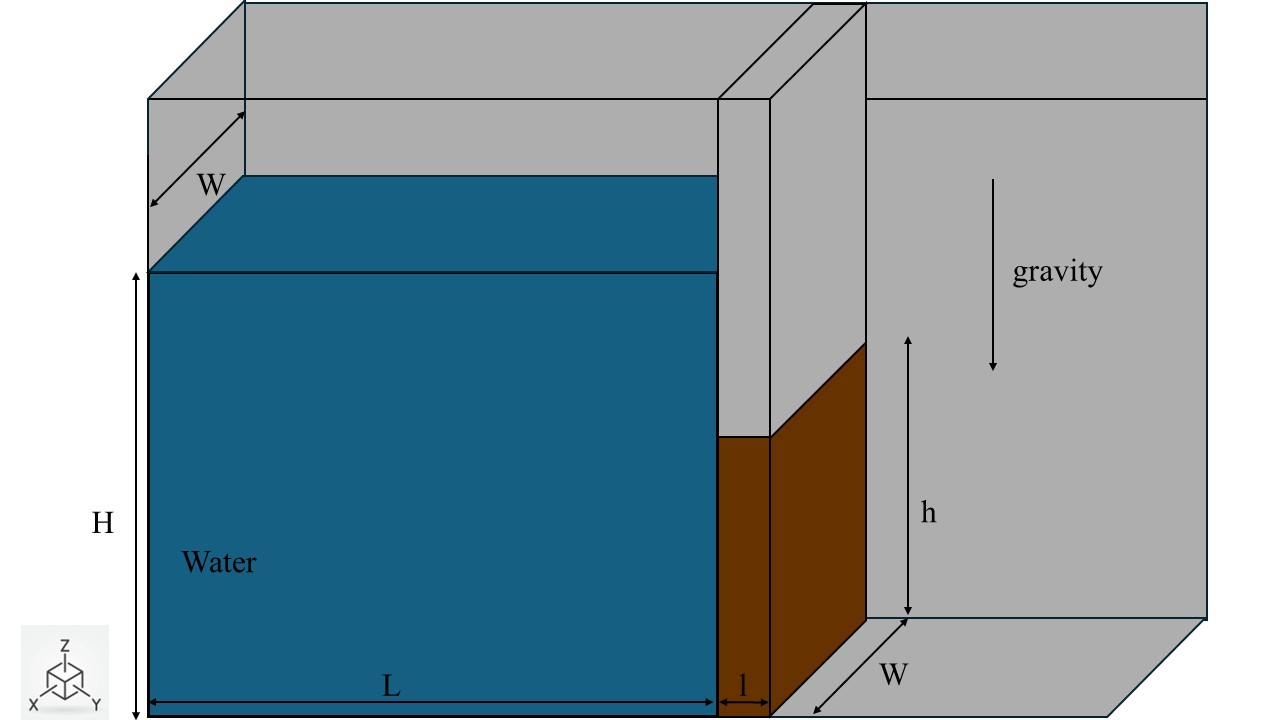}
       \caption{Illustration of the setup for a dam-break simulation involving an elastic gate, showing the configuration of the dam, the initial water height, and the position of the elastic gate.}
       \label{Confined water}
\end{figure}

 \begin{figure}[htb!]
    \centering
    \begin{subfigure}[b]{0.48\textwidth}
        \centering
        \includegraphics[width=\textwidth]{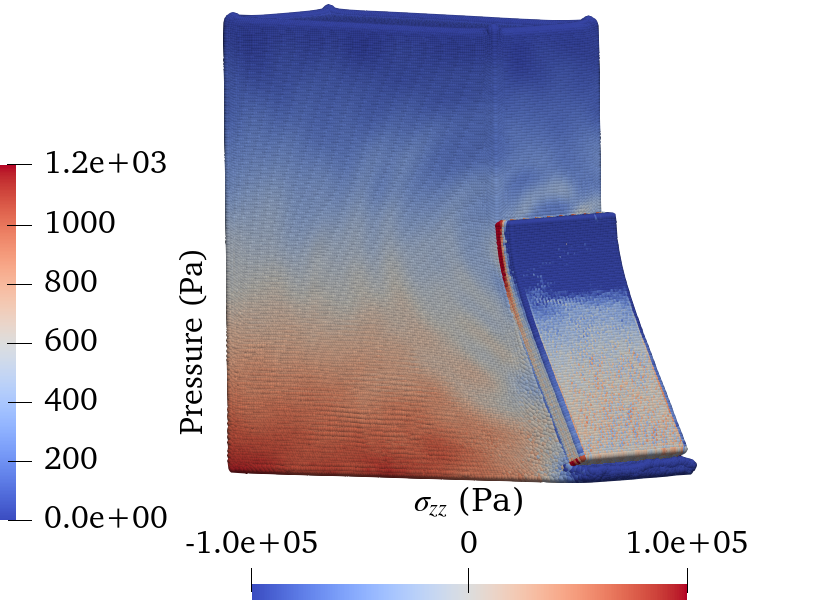}
        \caption{Time = 0.06s}
        \label{fig:time006}
    \end{subfigure}
    \hfill
    \begin{subfigure}[b]{0.48\textwidth}
        \centering
        \includegraphics[width=\textwidth]{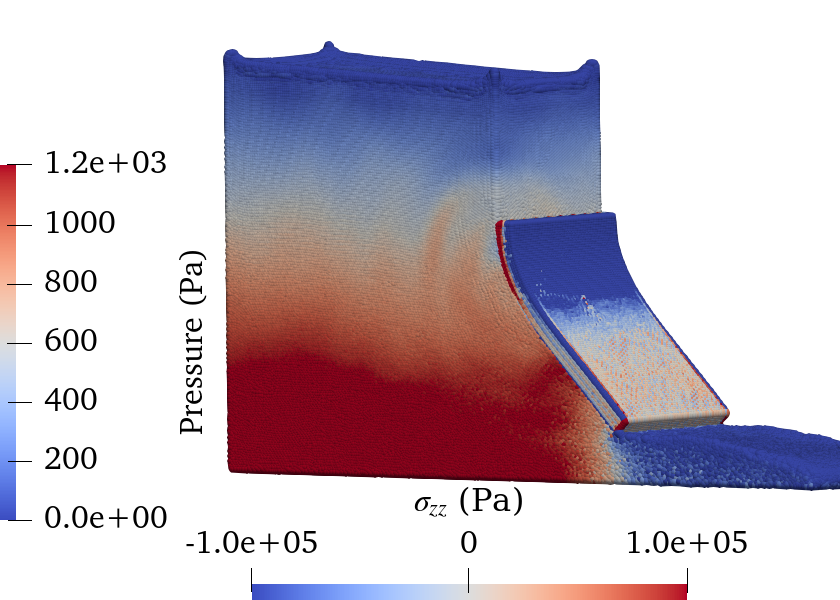}
        \caption{Time = 0.12s}
        \label{fig:time012}
    \end{subfigure}


    \begin{subfigure}[b]{0.48\textwidth}
        \centering
        \includegraphics[width=\textwidth]{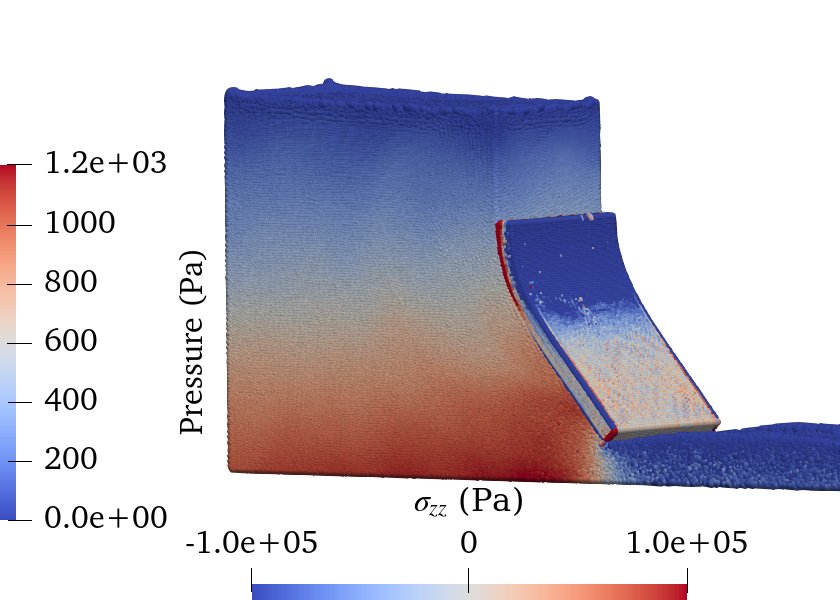}
        \caption{Time = 0.18s}
        \label{fig:time018}
    \end{subfigure}
    \hfill
    \begin{subfigure}[b]{0.48\textwidth}
        \centering
        \includegraphics[width=\textwidth]{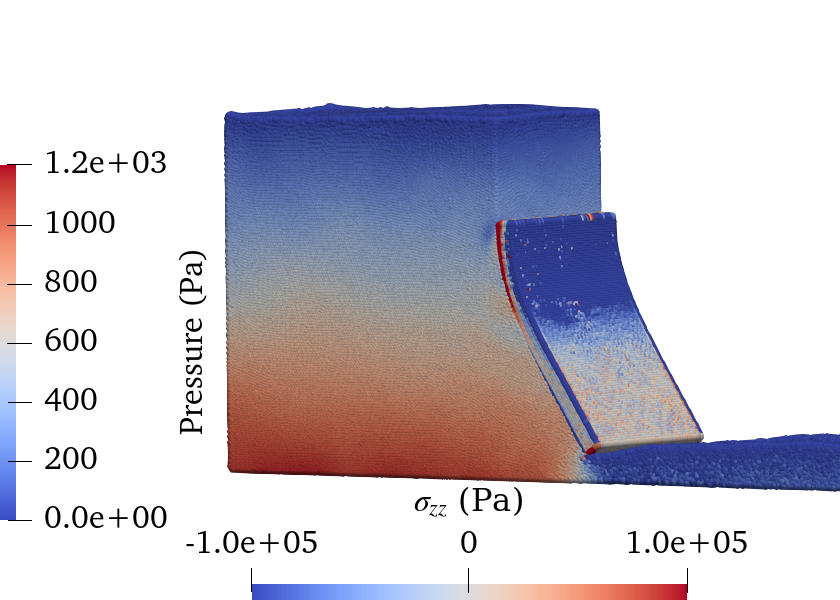}
        \caption{Time = 0.24s}
        \label{fig:time024}
    \end{subfigure}

    \caption{Contour plots at different time steps depict the water column's flow and pressure $p_r$ distribution, as well as the stress ($\sigma_{zz}$) distribution of the deformable elastic gate. Each plot shows the system's evolution over time, showcasing the dynamic behavior of the water and structure and their interactions.}
    \label{Water_flow}
\end{figure}

The setup is shown in Figure \ref{Confined water}. The setup contains three parts - water, a rigid wall, and a deformable gate fixed at its top end. The height and length of the water columns are $H=0.14$~m  and $L=0.1$~m, respectively. The rubber gate's vertical length ($h$) is $0.079$~m, and the thickness (l) is $0.005$~m. The width of the entire setup is taken as $W=0.05$~m. The densities of water and the rubber wall are 1000 kg/m$^3$ and 1100 kg/m$^3$, respectively. For the rubber, Young's modulus ($E$) is $12$ MPa and Poisson's ratio ($\nu$) is $0.45$. The test case is simulated with a time-step of $\Delta t = 5 \times 10^{-6}$~s, and the initial computational domain is discretized with $\Delta p = 0.0008$~m and the smoothing length is $h=1.3\Delta p$. The artificial viscosity parameters ($\beta_1$ and $\beta_2$) are set as 0.03 and 0 for water, and 1.5 and 1.5 for the rubber gate. We have used $\gamma = 0.25$ to calculate the artificial pressure to suppress tensile instability. The contact between the elastic rubber gate and the water is modeled through boundary conditions discussed in the previous section. The interaction between water and the rigid wall is modeled using the boundary condition described in Section \ref{interaction}.

Initially, the deformable gate confines the water as shown in Figure \ref{Confined water}. The water is at rest and has a hydrostatic pressure distribution. When the experiment starts, the water starts to flow and push the gate to the right side, and exits the tank through the gap between the gate and the bottom of the tank. The flow process is shown in  Figure~\ref{Water_flow}. 

During the early stages, the horizontal displacement of the elastic gate increases quickly from 0.06 s to 0.12 s as it behaves like a cantilever rubber beam. The stress $\sigma_{zz}$ distributions in the rubber gate instances are also shown in Figure~\ref{Water_flow}. It is observed that the upstream face of the gate is in tension, whereas the external face is compressed. The maximum value of stress is found near the upper end of the gate, while the value becomes zero at the free end of the gate. The maximum value of stress and its distribution increase at the same time and then slightly decrease from 0.18 s to 0.24 s due to the decreasing water depth, and consequently, the pressure on the elastic gate diminishes, causing the gate to return to its initial position slowly. The presence of noise in the stress distribution of the elastic gate can be further reduced by using particle shifting \cite{xu2018technique}, stress point method \cite{dyka1997stress}, hour-glass control scheme \cite{mohseni2021application, shimizu2022implicit}, Shepard filter \cite{mayrhofer2013investigation}. However, the implementation of these numerical schemes is beyond the scope of the present work.

\begin{figure}[htb!]
    \centering
    \begin{subfigure}[b]{1\textwidth} 
        \centering
        \includegraphics[width=\textwidth]{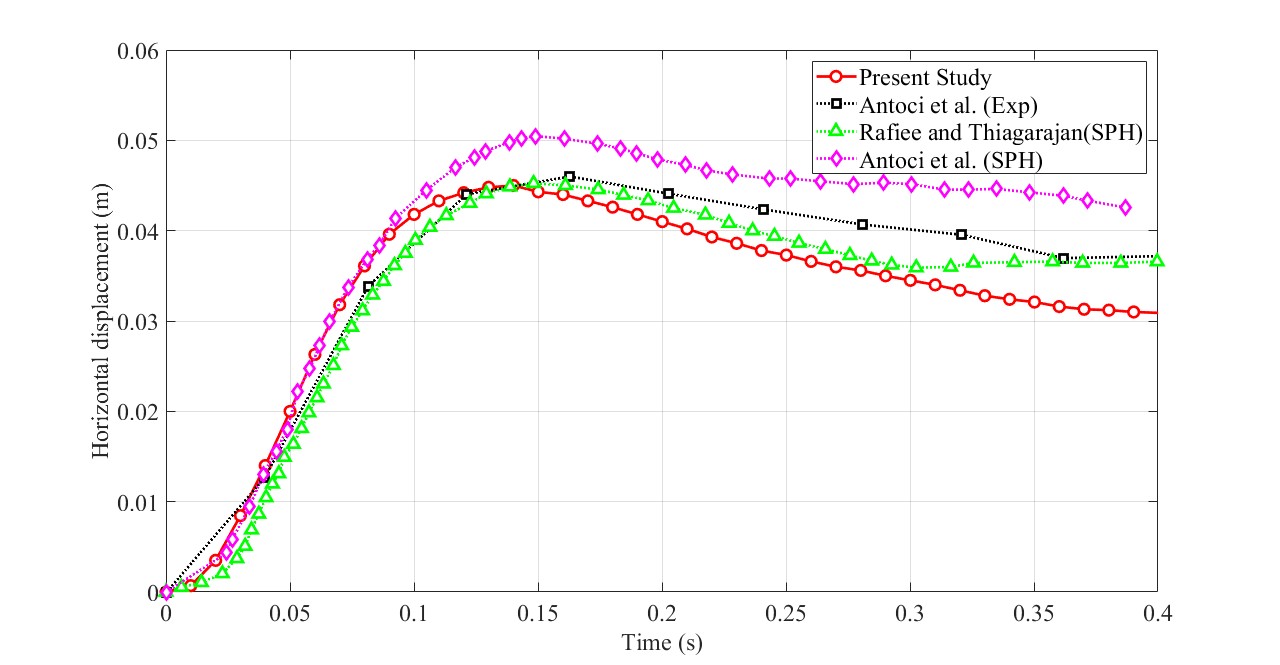}
        \caption{Horizontal displacement}
    \end{subfigure}
    \hfill
    \begin{subfigure}[b]{1\textwidth} 
        \centering
        \includegraphics[width=\textwidth]{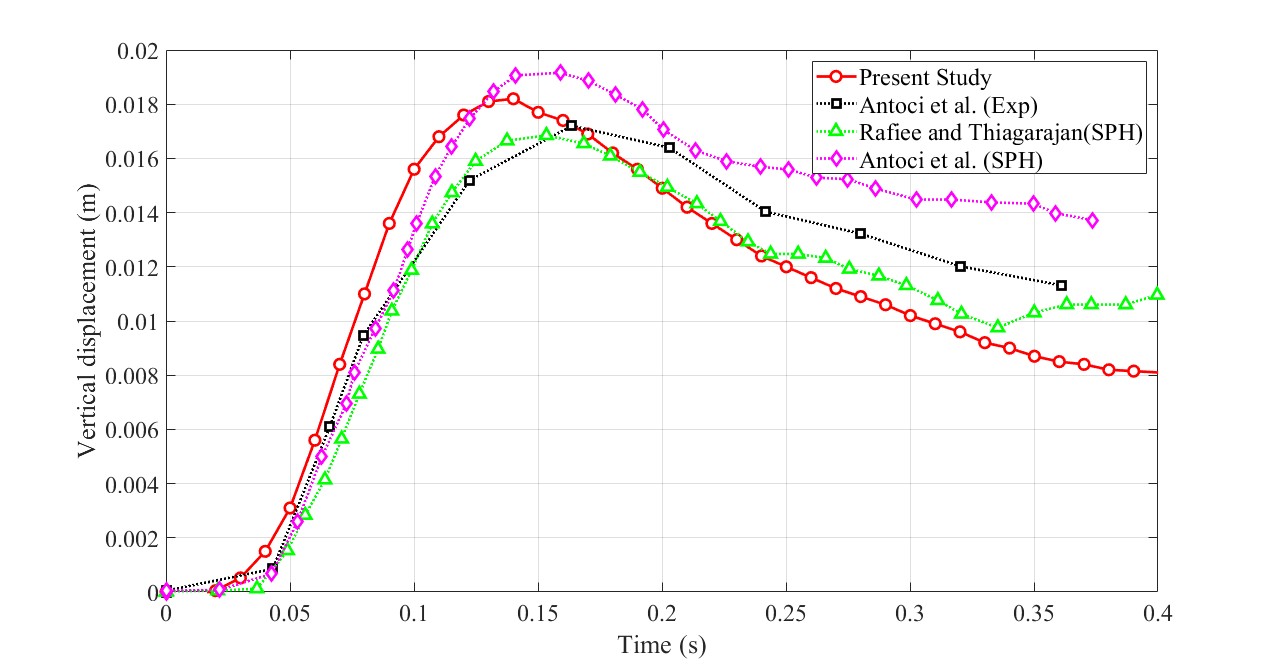}
        \caption{Vertical displacement}
    \end{subfigure}
    \caption{Time-history plot showing the rubber gate's displacement at its free end and comparing the simulation results with experimental data and numerical results from the literature.}
    \label{Horizontal_Vertical}
\end{figure}

To validate the developed method quantitatively, the history of the displacements at the free end of the gate is shown in Figure~\ref{Horizontal_Vertical}, along with experimental data and other reference numerical results. The present prediction of horizontal and vertical displacement of the rubber gate is comparable with the other results available in the literature \cite{antoci2007numerical, rafiee2009sph}. Moreover, The simulation remains stable throughout, with no instability or breakdown encountered. The results show that the structure solver with intermediate neighbors and corrected kernel gradient performs well and can be used to model the structure response under hydrodynamic loading.

\subsection{Breaking dam flow impact on flexible obstacle}

\begin{figure}[htb!]
    \centering
    \includegraphics[width=1\textwidth]{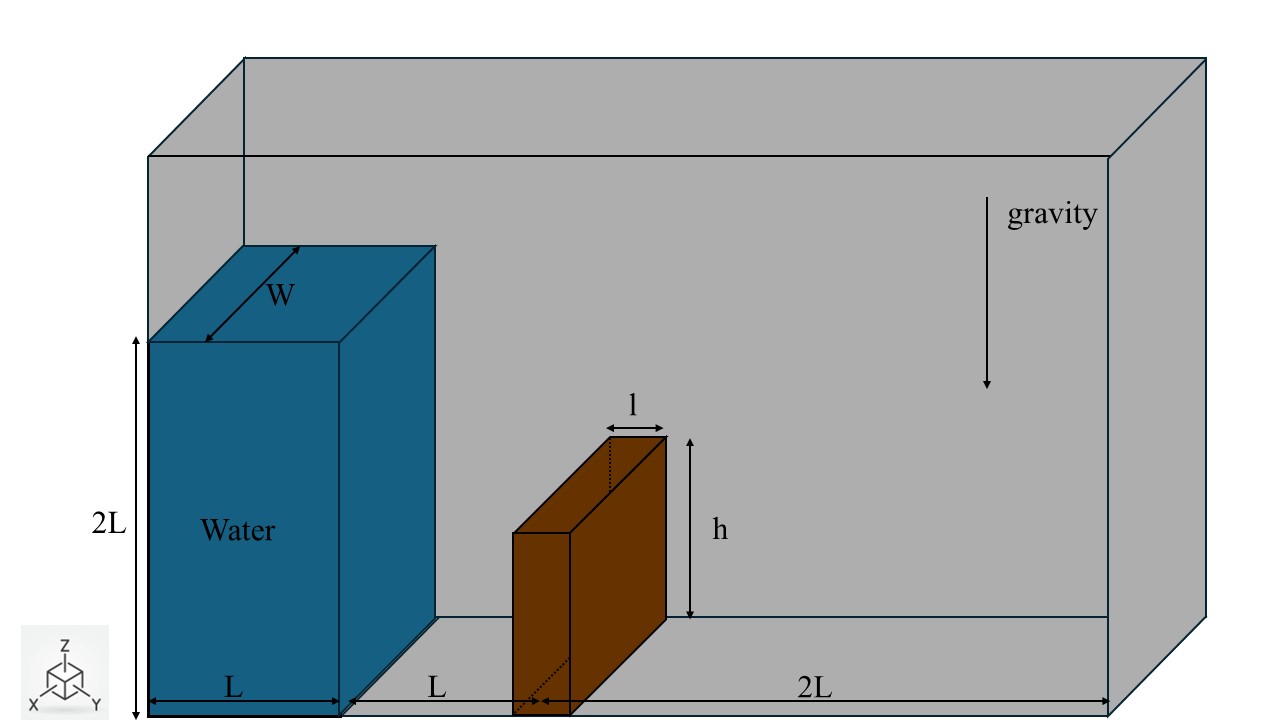}
    \caption{Initial system configuration, depicting the breaking dam flow impacting an elastic wall. The no-slip boundary condition is imposed at the fluid-solid wall interface no-slip conditions.}
    \label{obstacle}
\end{figure}

In this example, we simulate the impact of water flow onto an elastic object at the midpoint of a tank, as shown in Figure \ref{obstacle}. Initially, water is stored in a confined area (Length $L=0.146$~m; Height $H=0.292$~m; Width $W=0.05$~m), and the elastic obstacle (Height $h=0.08$~m; Width $W=0.05$~m; Thickness $l=0.012$~m) is located at the mid-point ($2L=0.292$~m). The bottom of the elastic obstacle is fixed, and the top end is free. The length of the tank is $4L=0.584$~m. The density of water is taken as $1000~Kg/m^3$. The density and elastic modulus of the elastic wall are taken as 2500~kg/m$^3$ and $10^6 $~N/m$^2$, respectively. The Poisson's ratio is kept at $0.0$, which is also used in the reference numerical simulations \cite{rafiee2009sph}. The inter-particle spacing is $\Delta p = 0.0025$~m and $h= 1.3\Delta p$. The time step for the integration is $\Delta t = 5 \times 10^{-6}$~s, with $\beta_1$ and $\beta_2$ set as 0.03 and 0.0 for water, and 1.5 and 1.5 for the deformable structure, respectively. We have used $\gamma = 0.25$ to calculate the artificial pressure to suppress tensile instability. Again, in this case, the structure failure is not considered.

\begin{figure}[htbp]
    \centering
    \begin{subfigure}[b]{0.48\textwidth}
        \centering
        \includegraphics[width=\textwidth]{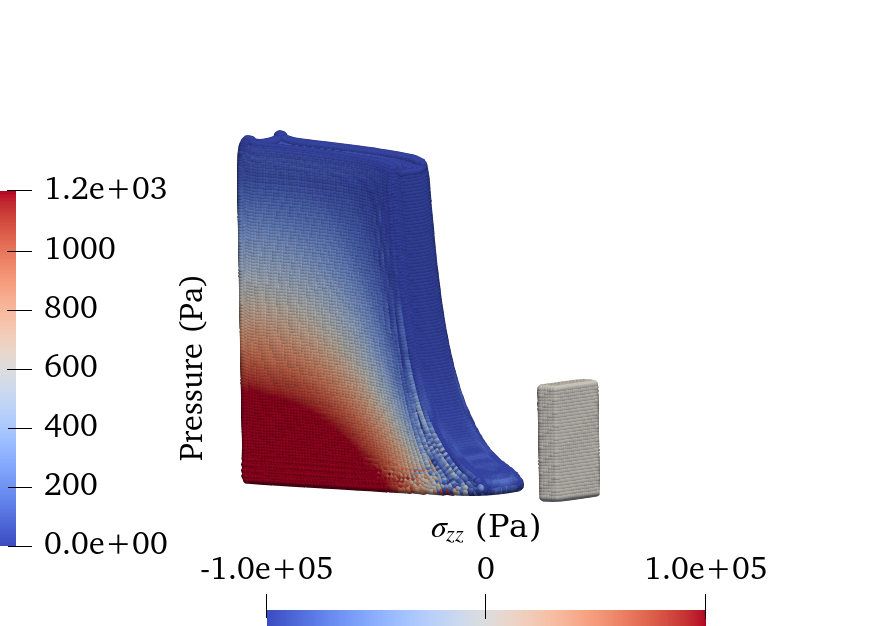}
        \caption{Time = 0.1s}
        \label{fig:time006}
    \end{subfigure}
    \hfill
    \begin{subfigure}[b]{0.48\textwidth}
        \centering
        \includegraphics[width=\textwidth]{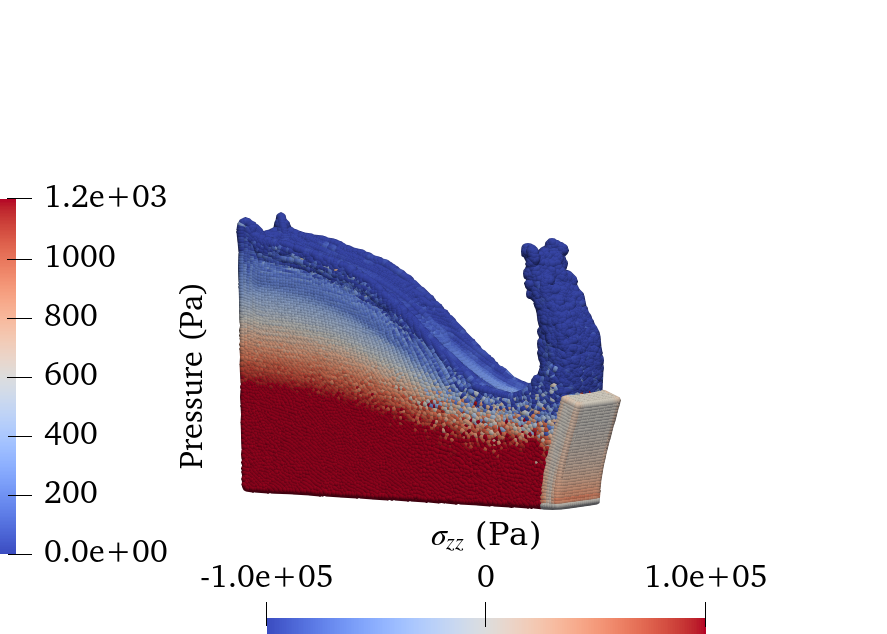}
        \caption{Time = 0.2s}
        \label{fig:time012}
    \end{subfigure}

    \vskip\baselineskip 

    \begin{subfigure}[b]{0.48\textwidth}
        \centering
        \includegraphics[width=\textwidth]{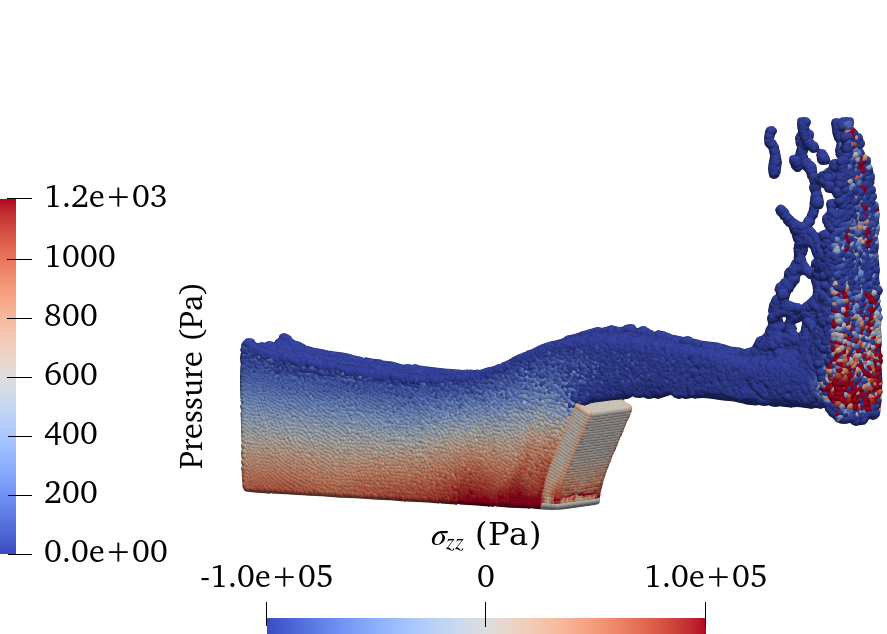}
        \caption{Time = 0.4s}
        \label{fig:time018}
    \end{subfigure}
    \hfill
    \begin{subfigure}[b]{0.48\textwidth}
        \centering
        \includegraphics[width=\textwidth]{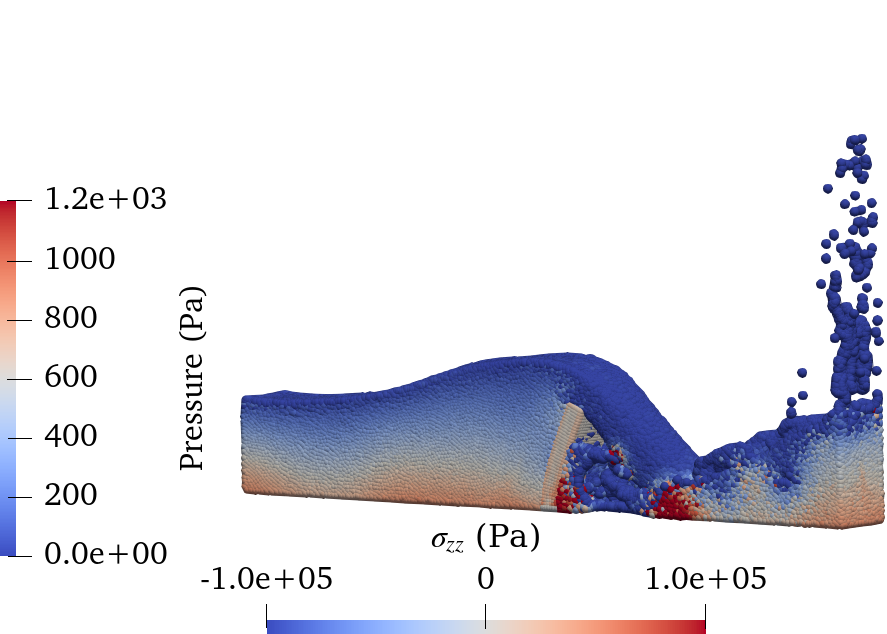}
        \caption{Time = 0.6s}
        \label{fig:time024} 
    \end{subfigure}
 \begin{subfigure}[b]{0.48\textwidth}
        \centering
        \includegraphics[width=\textwidth]{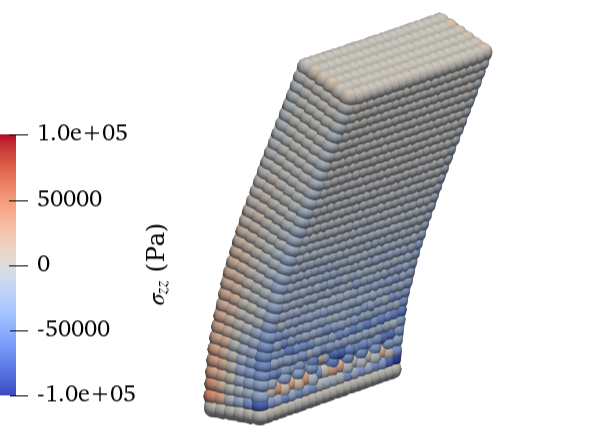}
        \caption{Time = 0.6s}
        \label{fig:time024} 
 \end{subfigure}

\caption{Pressure and stress distributions at various time steps during the water impact on an elastic obstacle, illustrating how the stress varies across the obstacle's surface as the fluid interacts.}
    \label{both}
\end{figure}

This simulation provides a detailed representation of different features in problems of hydrodynamically loaded structures, including quick changes in water pressure, violent free surfaces, and large deformation structures (Figure \ref{both}). At first, the water moves without obstacles, and the flow's front edge shows a low-pressure level. Nevertheless, as the water encounters the flexible barrier, a notable increase in pressure becomes evident, resulting in high-impact forces. As a result, there is deformation in the flexible barrier, which leads to noticeable alterations in the movement of the flowing water. As the water flows over the wall, the pressure decreases progressively. At this point, the elastic wall starts moving towards its original shape after being distorted. Once the water hits the tank's solid wall at the far end, a concentrated area of high pressure is formed again. In the deformable obstacle, the highest level of tension is located at the immovable extremity of the flexible barrier, whereas it is absent at the unrestricted extremity.

\begin{figure}[htbp]
    \centering
    \includegraphics[width=1\textwidth]{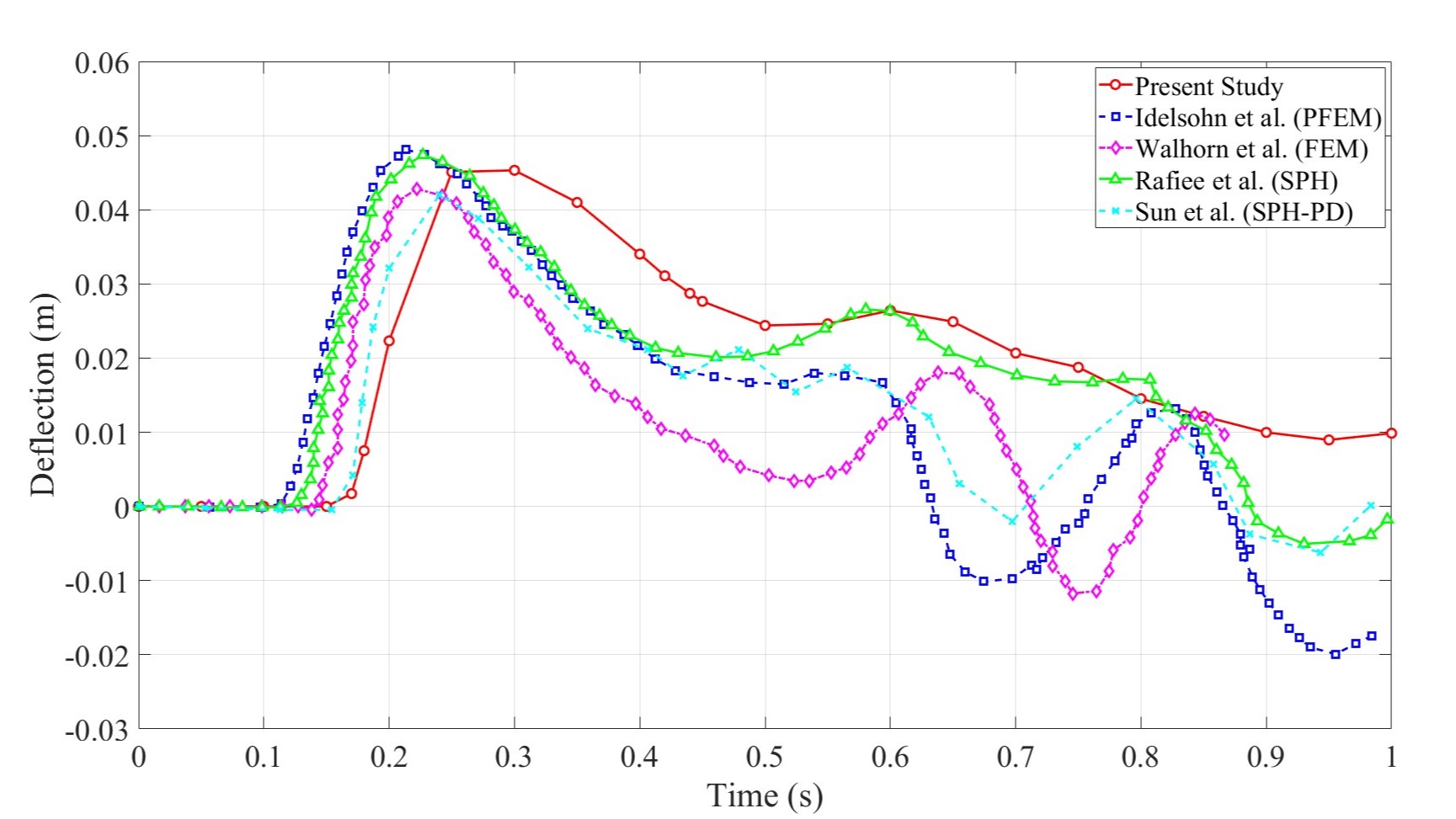} 
    \caption{Time-history plot showing the horizontal deflection of the upper-left corner of the elastic wall, tracking its displacement over time in response to fluid impact. This also compares the simulated deflection with other results from the literature to assess the accuracy and response of the present model.}
    \label{deflection}
\end{figure}

To validate the effectiveness of our method, we examined the deflection process at the upper-left corner of the elastic plate over time \cite{rafiee2009sph, zhan2019stabilized, rahimi2023sph}. Figure \ref{deflection} illustrates the variations in horizontal displacement measured at this corner. Additionally, we include quantitative data from other studies for comparative analysis \cite{rafiee2009sph, sun2020smoothed, walhorn2005fluid, idelsohn2008unified}. The results indicate that our proposed method reliably predicts and replicates the overall response of the obstruction when subjected to hydrodynamic forces generated by the collapsing water column.

\subsection{Fracture of a brittle structure under the impact of dam break flow}
In this example, we focus on modeling the damage propagation leading to the failure of the elastic obstacle due to the hydrodynamic impact loading from the previous example in Section~5.4. The cracking is assumed to be brittle, having a maximum fracture strain of $\epsilon_f^{\max}$ = 0.09, similar to the assumption in \cite{dai2023coupled}. Therefore, the interaction coefficient $f_{ij}$ is 1 (full interaction) if $\epsilon_f<0.09$, while $f_{ij} = 0$ (no interaction) if $\epsilon_f~\ge$ 0.09, which signifying development of cracks between the connecting particle pair $i$ and $j$. The SPH and material parameters are kept unchanged from the last example.

\begin{figure}[htbp]
    \centering
    \begin{subfigure}[b]{0.48\textwidth}
        \centering
        \includegraphics[width=\textwidth]{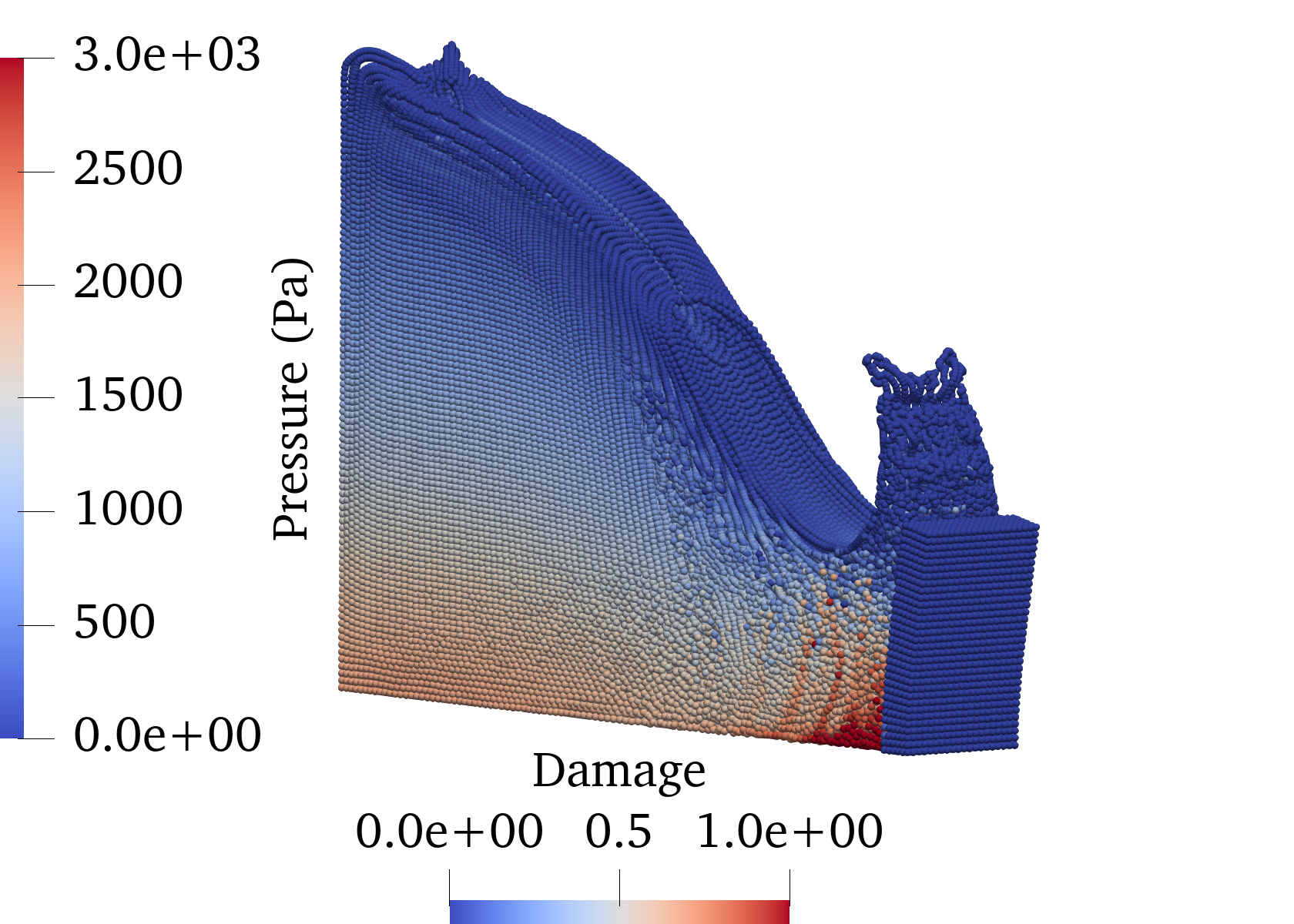}
        \caption{Time = 0.2s}
        \label{fig:time006}
    \end{subfigure}
    \hfill
    \begin{subfigure}[b]{0.48\textwidth}
        \centering
        \includegraphics[width=\textwidth]{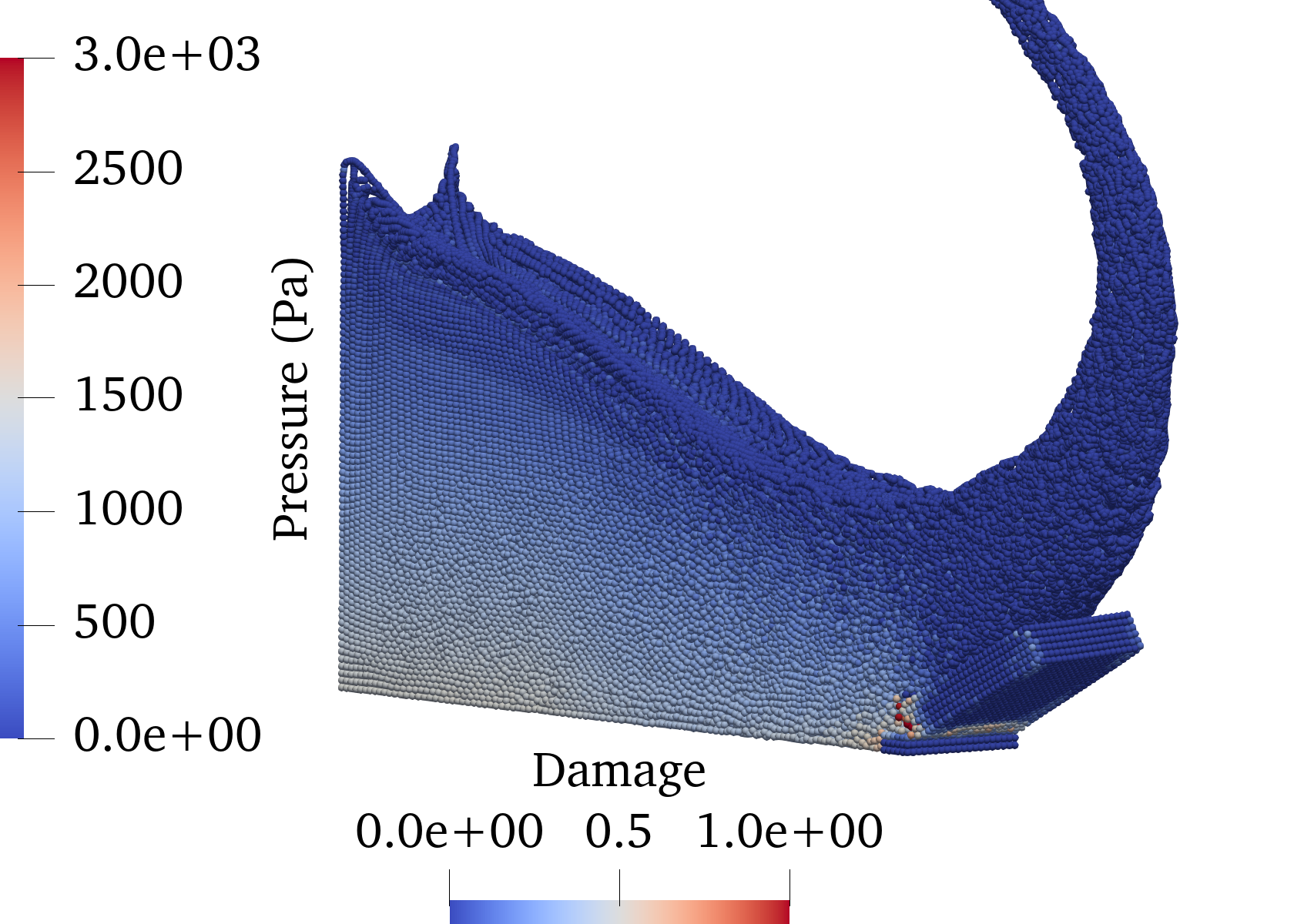}
        \caption{Time = 0.3s}
        \label{fig:time012}
    \end{subfigure}

    \vskip\baselineskip 

    \begin{subfigure}[b]{0.62\textwidth}
        \centering
        \includegraphics[width=\textwidth]{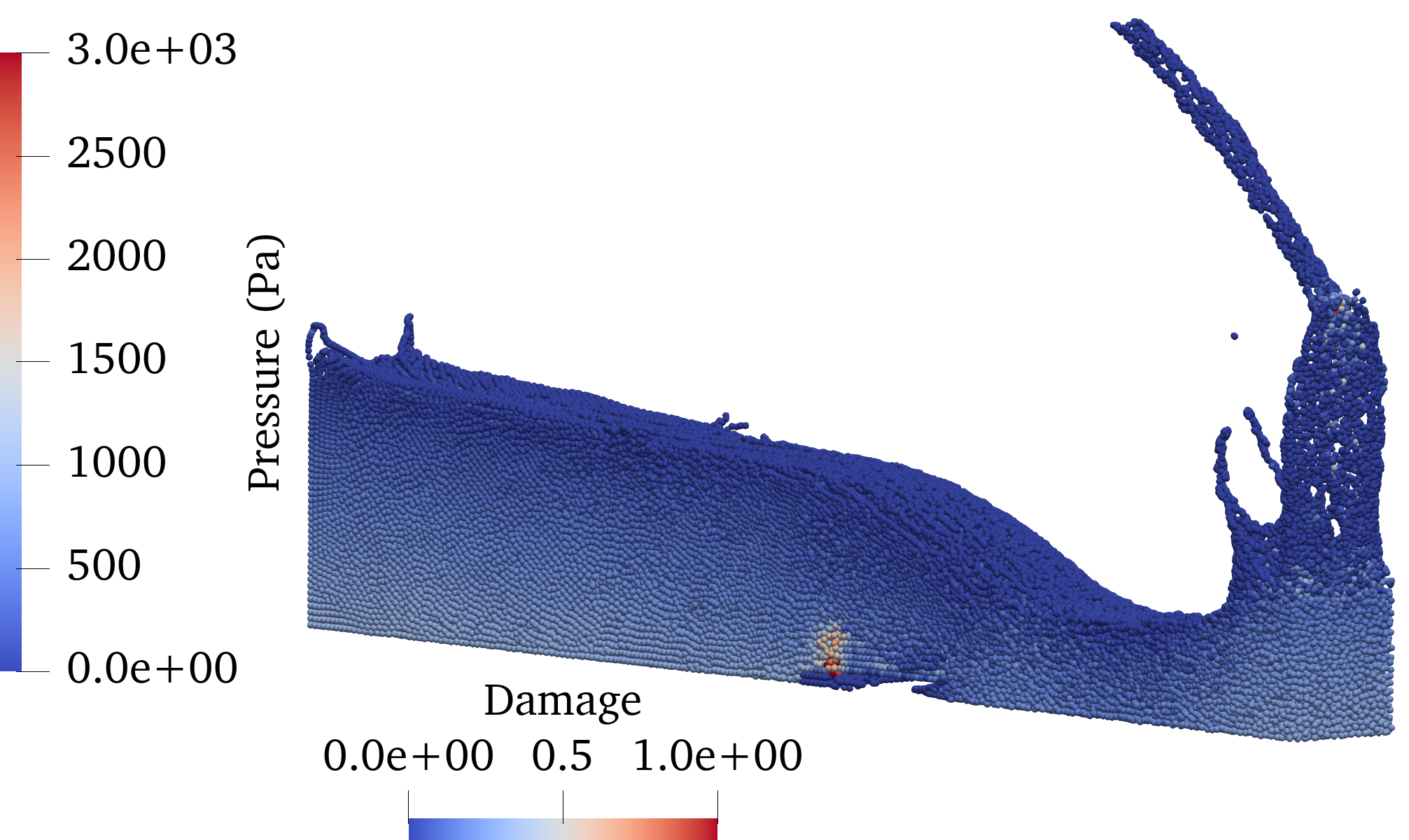}
        \caption{Time = 0.5s}
        \label{fig:time018}
    \end{subfigure}

\caption{Pressure and stress distributions at various time steps during the water impact on an elastic obstacle, illustrating how the stress varies across the obstacle's surface as the fluid interacts.}
    \label{both}
\end{figure}

\begin{figure}[htbp]
    \centering
    \includegraphics[width=1\textwidth]{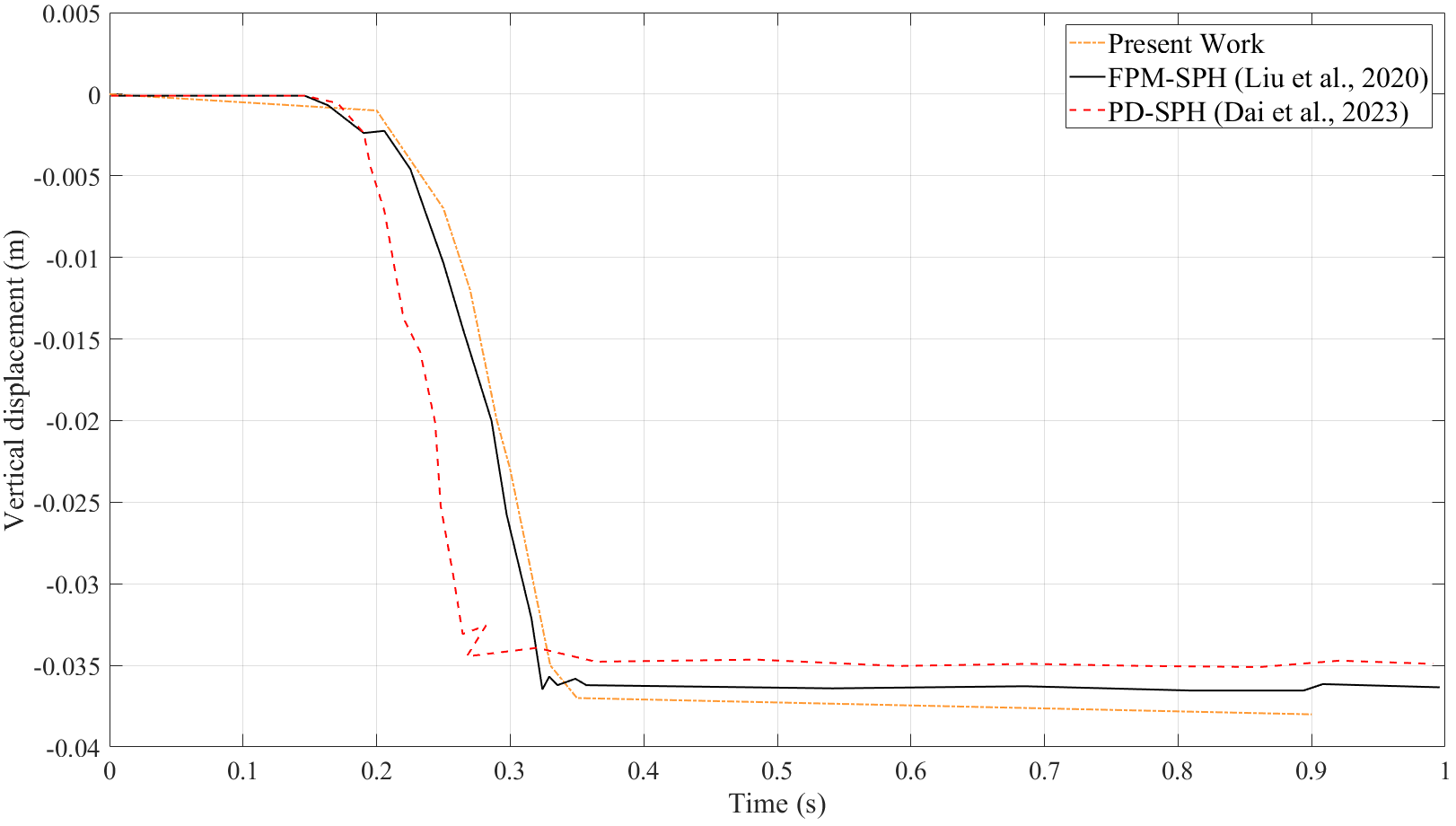} 
    \caption{Displacement time history of the deformable elastic obstacle at the midpoint and its comparison with other results from the literature.}
    \label{frac1}
\end{figure}

Figure \ref{both} displays a sequence of simulation snapshots, showing the progression of a dam-break event and the resulting brittle fracture in the baffle. These visuals capture the key stages of the failure process. The pressure within the water and the damage pattern within the baffle are also depicted in Figure \ref{both}. Figure \ref{frac1}, on the other hand, illustrates the vertical displacement at the centre of the baffle. Around t = 0.2 s, material damage initiates near the baffle's fixed end, followed by the propagation of this damage, which is accompanied by oscillations in the displacement, leading to instability. A crack begins to form at t = 0.22 s, and by t = 0.25 s, the baffle breaks apart, with fragments being swept downstream and reaching the bottom at approximately t = 0.3 s. The displacement evolution at the midpoint is compared to the results from the FPM-SPH \cite{liu2020coupled} and SPH-Peridynamics \cite{dai2023coupled} methods. The proposed model in this study aligns closely with the FPM-SPH and SPH-Peridynamics results, particularly in terms of the timing of damage onset and the final failure, demonstrating the accuracy of this framework for fracture analysis in fluid-structure interactions.

\begin{figure}[htbp]
    \centering
    \includegraphics[width=1\textwidth]{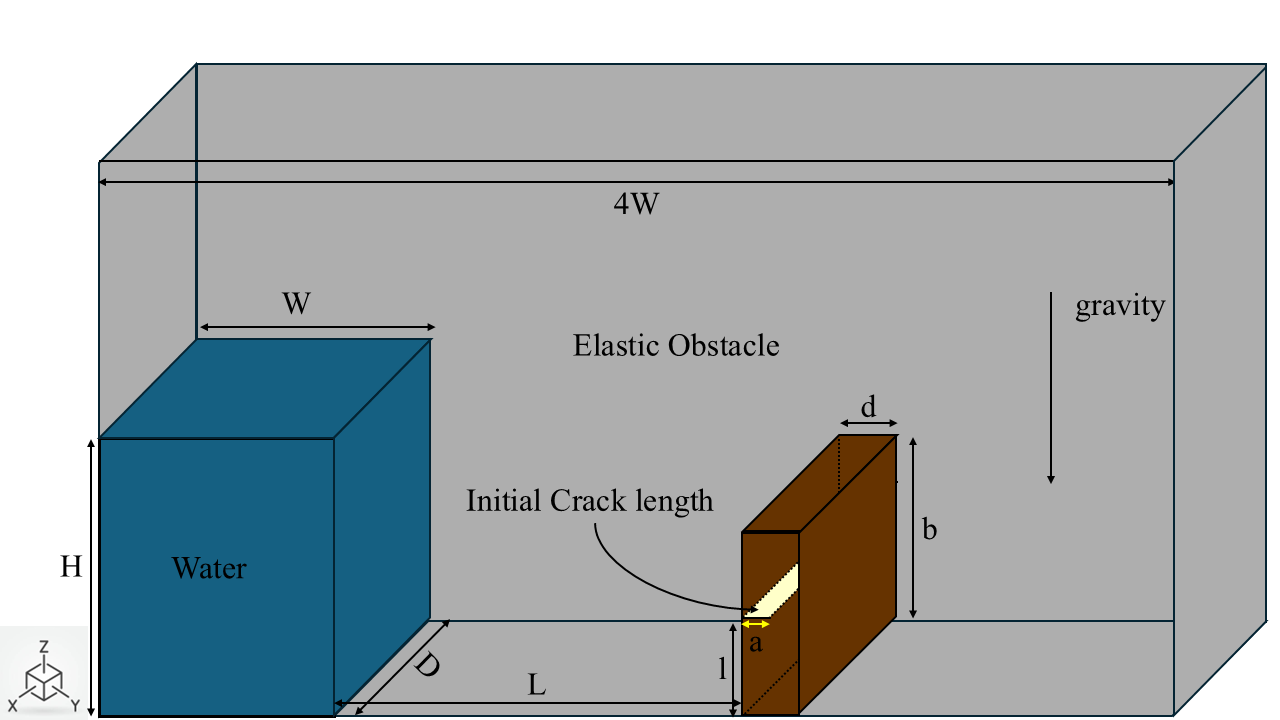} 
    \caption{Initial system configuration, depicting the breaking dam flow impacting a deformable elastic obstacle with a notch.}
    \label{frac3}
\end{figure}

\subsection{Fracture of pre-notched elastic obstacle subjected to water impact}
We investigate the interaction between water flow and a brittle obstacle with a notch. The geometric configuration for this scenario is outlined in Figure \ref{frac3} \cite{rahimi2023sph}. In this case, the notch is made by removing a line of particles. The obstacle, which includes a pre-existing crack of length $a = 0.008$ m, is situated $l = 0.025$ m above the bottom. The system dimensions are: $H = 0.3$ m, $W = 0.15$ m, $L = W$, $b = 0.09$ m, and $d = 0.03$ m. The process begins with the abrupt release of the water body confined initially on the left, then the water flows towards the obstacle and impacts with it. The water and obstacle have respective densities of 1000 kg/m$^3$ and 800 kg/m$^3$ \cite{rahimi2023sph}, and the obstacle's material properties are: elastic modulus of $E = 9.7 \times 10^6$ N/m² and a Poisson ratio of $\nu = 0.17$. The maximum fracture strain, $\epsilon_f$, is set at 0.01. When the strain between any two particles $i$ and $j$ exceeds this value, their connection is severed, permanently halting the interaction ($f_{ij} = 0$ when $\epsilon_f > \epsilon_f^{\max}$). The fracture is considered irreversible throughout the simulation. The spatial resolution used is $\Delta p = 0.0025$ m, with $h/\Delta p = 1.3$, and the time step for integration is $\Delta t = 1~\mu$s. For water, the parameters $\beta_1$ and $\beta_2$ are set to 0.1, while for the elastic structure, they are 1.5. We have used $\gamma = 0.9$ to calculate the artificial pressure to suppress tensile instability.

\begin{figure}[htbp]
    \centering
    \begin{subfigure}[b]{0.48\textwidth}
        \centering
        \includegraphics[width=\textwidth]{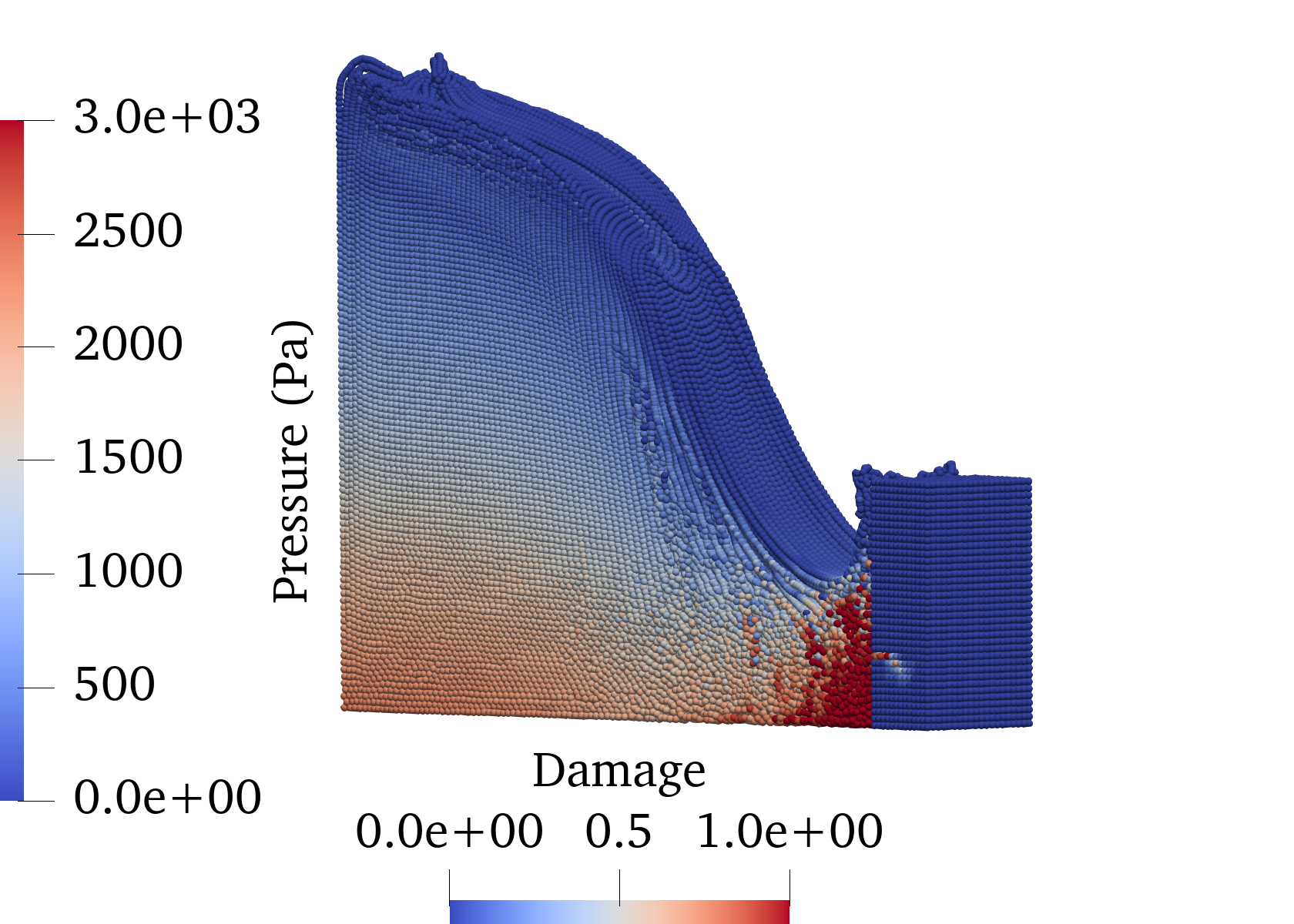}
        \caption{Time = 0.18s}
        \label{fig:time006}
    \end{subfigure}
    \hfill
    \begin{subfigure}[b]{0.48\textwidth}
        \centering
        \includegraphics[width=\textwidth]{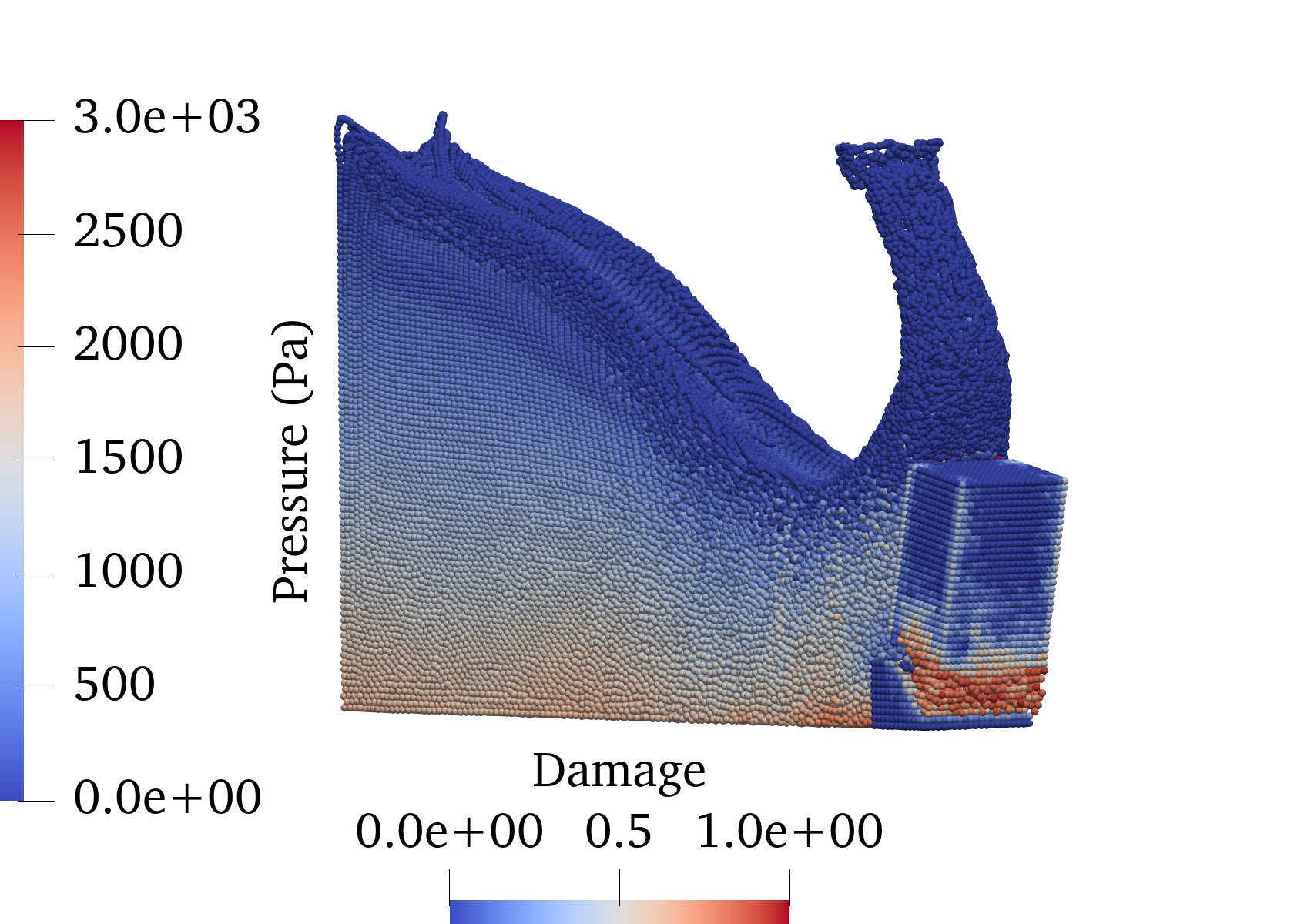}
        \caption{Time = 0.23s}
        \label{fig:time012}
    \end{subfigure}
    \caption{Illustration of the damage and failure progression of an elastic obstacle at different time steps, showcasing how the material of the obstacle deforms and fractures under fluid impact.}
    \label{frac4}
\end{figure}

\begin{figure}[htb!]
    \centering
    \includegraphics[width=0.7\textwidth]{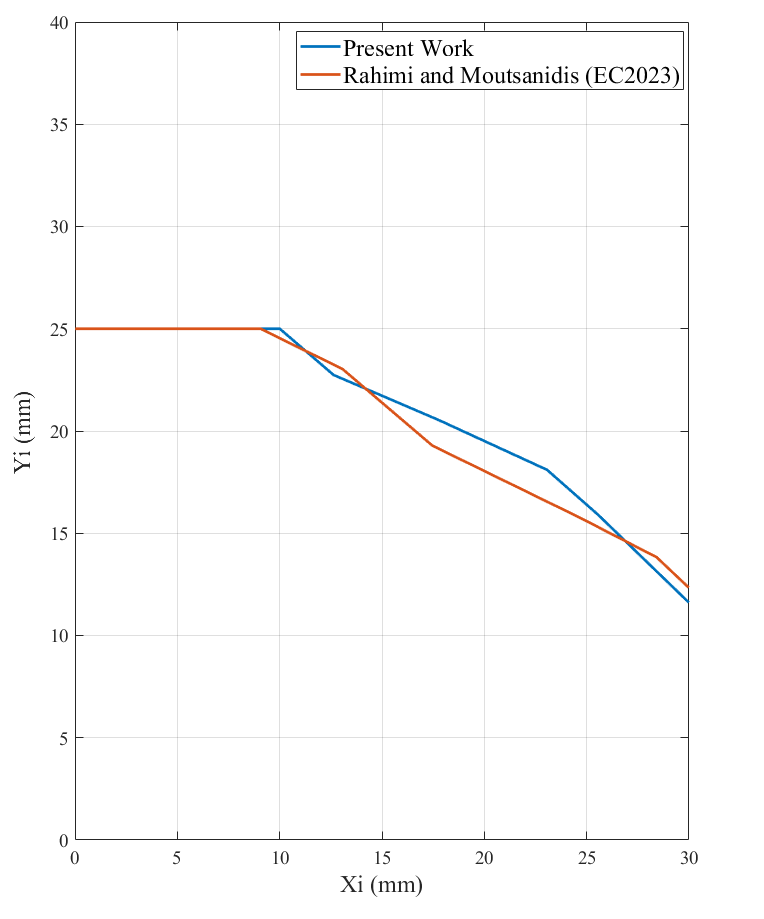} 
    \caption{Crack or fracture pattern in the deformable structure predicted using the present framework and its comparison with prior research findings \cite{rahimi2023sph}. The horizontal line denotes the initial notch length.}
    \label{frac5}
\end{figure}

Figure \ref{frac4} illustrates the changes in the water surface, pressure distribution, and crack growth as the simulation progresses. Initially, as the fluid impacts the obstacle, pressure in the fluid-structure interaction region sharply rises, causing the obstacle to deform. As the strain accumulates over time, it exceeds the fracture limit ($\epsilon_f^{\max} = 0.01$) at around 0.162 s, triggering crack growth and separating the material at the crack tip. By 0.185 s, the upper portion of the obstacle detaches, while the lower section produces a jet-like effect, guiding the water along a specific path. The last snapshots in Figure \ref{frac4} capture this process. Similar patterns, such as crack initiation at 0.165 s and full detachment at 0.188 s, have been observed in other studies. Figure \ref{frac5} shows that the final fracture pattern is consistent with prior research findings \cite{rahimi2023sph}.

\section{Conclusion}\label{conclu}
This research develops a 3D SPH model to simulate fluid-structure interaction considering structure damage and failure. The approach employs two key strategies: for the fluid, a weakly compressible SPH (WCSPH) method with a density diffusion term improves accuracy, while boundary interactions with rigid surfaces are managed using boundary particles. A pseudo-spring SPH approach is adopted for the deformable solid, allowing material damage and crack propagation to be modeled efficiently, avoiding complex methods like visibility criteria or particle splitting.

A modified boundary condition is applied for fluid-solid interactions, eliminating the need for penalty force calibration. In this method, the interaction between water and elastic structure particles is captured by representing the structure particles as dynamic boundaries influencing nearby fluid/water particles. The fluid/water pressure is extrapolated onto the structure particles to evaluate the interaction forces, which are then applied symmetrically following Newton’s third law. The results are validated through comparison with analytical solutions, experiments, and other numerical methods, showing that the framework accurately captures free surface flow and structure dynamic responses without instability. Several case studies on fluid-structure interaction (FSI) problems involving damage and fracture in deformable bodies confirm the method’s reliability, with results closely matching existing literature.

While the method shows promising accuracy, further validation through real-world experiments, particularly in structural and fracture cases, is necessary. Future work will focus on laboratory experiments to refine the approach. Enhancements like tensile instability control and particle shifting could improve performance.

\section{Acknowledgments}
MRII acknowledges the computational support provided as a part of the SRG DST-SERB (Grant No. SRG/2023/001266) and the IIT Delhi, India NFS grant on which the simulations have been run.

\FloatBarrier



\end{document}